\documentclass[11pt]{article}
\usepackage[utf8]{inputenc}
\usepackage[T1]{fontenc}

\usepackage{jheppub}

\usepackage{amsfonts}
\usepackage{amsthm}
\usepackage{mathtools}
\usepackage{slashed}
\usepackage{tensor}
\usepackage{physics}

\usepackage{float}
\usepackage{array}

\usepackage{graphicx}
\usepackage[dvipsnames]{xcolor}
\usepackage{enumitem}
\usepackage{lipsum}
\usepackage{subfigure}

\usepackage{hyperref}

\newcommand{\ads}{\ensuremath{\mathrm{AdS}}}
\newcommand{\adsf}{\ensuremath{\mathrm{AdS}_5}}

\newcommand{\Nfour}{\ensuremath{\mathcal{N}=4}}

\newcommand{\halfBPS}{1/2-BPS}
\newcommand{\quartBPS}{1/4-BPS}

\newcommand{\Op}{\ensuremath{O}}
\newcommand{\bigO}{\ensuremath{\mathcal{O}}}
\newcommand{\vol}{\ensuremath{\text{vol}}}

\newcommand{\normCorr}[1]{\ensuremath{\langle\!\langle\, #1 \,\rangle\!\rangle}}

\DeclareMathOperator*{\diag}{diag}

\begin{document}
	\title{Holography of quarter-BPS AdS bubbles}

    \author[a]{Bogdan Ganchev}
    \affiliation[a]{William H. Miller III Department of Physics and Astronomy, \\
Johns Hopkins University, 3400 North Charles Street, Baltimore, Maryland, 21218, USA}
    \emailAdd{bganche1@jh.edu}
    
	\author[b]{Anthony Houppe}
	\affiliation[b]{Institut für Theoretische Physik, ETH Zürich, \\ Wolfgang-Pauli-Strasse 27, 8093 Zürich, Switzerland}
	\emailAdd{ahouppe@phys.ethz.ch}

	\abstract{
    We consider the quarter-BPS sector of the AdS$_5$/CFT$_4$ duality, and provide a precise matching between specific CFT states and supergravity geometries beyond the linearized approximation. In the bulk, we focus on AdS bubbles geometries that fit within a consistent truncation to five-dimensional gauged supergravity. 
    We show that the BPS equations can be decoupled, and study the resulting geometries both perturbatively and numerically. 
    Applying the holographic dictionary, we compute the expectation values of light chiral primaries in these backgrounds. This data allows us to determine the dual CFT state up to quadratic order in the fluctuations, and to provide non-trivial consistency checks of the result.
    The resulting state is expressed as a linear combination of half-BPS double-trace operators, and of the unique $[2,0,2]$ quarter-BPS operator of the theory. As an application, we compute three-point functions involving two quarter-BPS operators and a half-BPS operator in the supergravity regime, finding an explicit example of a correlator that differs from its free theory value and that is therefore not protected.
    }

	\maketitle
	\flushbottom

	\section{Introduction}
	\label{sec:intro}

    The AdS/CFT correspondence is a powerful conjecture that posits a duality between a quantum theory of gravity and a non-gravitational conformal field theory living on its boundary \cite{Maldacena:1997re}. A key aspect of the holographic dictionary is that the bulk dual of a CFT state depends on how its conformal dimension scales with the central charge. \emph{Light} states, whose dimension does not scale with it, are dual to supergravity fluctuations (or string excitations) propagating on the AdS vacuum geometry. Conversely, \emph{heavy} (or \emph{huge}) states, with conformal dimension of order of the central charge, backreact on the geometry. A coherent heavy state is typically related to a smooth, horizonless classical solution of supergravity, asymptotic to \ads{}. When these states have the right charges, their dual is a black hole microstate, allowing one to study the quantum nature of black holes in different regimes.

    The D1-D5 CFT stands as the most successful realization of this idea. Initiated by the famous counting of Strominger and Vafa \cite{Strominger:1996sh}, the study of black hole microstates in that system has led to large families of states and their duals being matched on both sides of the duality \cite{Lunin:2001jy,Bena:2015bea,Bena:2016ypk,Bena:2017xbt,Ceplak:2018pws,Heidmann:2019xrd,Ganchev:2023sth,Ceplak:2024dbj} (see also \cite{Bena:2022rna,Bena:2025pcy} for recent reviews). This list includes smooth, horizonless geometries dual to 1/8-BPS states that contribute to the entropy of the D1-D5-P black hole. The identification process has been heavily aided by the general formalism of \cite{Skenderis:2006uy,Skenderis:2007yb,Kanitscheider:2006zf}, which allows one to compute protected three-point correlators to guide the identification between supergravity solutions and CFT states, as well as perform non-trivial tests of the dictionary \cite{Giusto:2015dfa,Giusto:2019qig,Rawash:2021pik,Turton:2025svk}. These precise matchings have further been used to compute Heavy-Heavy-Light-Light (HHLL) and Light-Light-Light-Light (LLLL) correlators in the strong coupling regime \cite{Galliani:2017jlg,Bombini:2017sge,Giusto:2018ovt,Bombini:2019vnc,Ceplak:2021wzz,Turton:2024afd}, by solving the wave equation in the supergravity background dual to the heavy states.

    Although the black hole entropy has been reproduced through microscopic state counting in other holographic systems \cite{Sen:2007qy,Benini:2015eyy,Azzurli:2017kxo,Cabo-Bizet:2018ehj}, the construction of actual black hole microstates and subsequent understanding of the holographic dictionary have had less success. 
    While \adsf{}/CFT$_4$ is one of most studied examples of the duality, the precise holographic dictionary is currently only well-understood in the \halfBPS{} sector. In Type IIB supergravity, a complete classification of the geometries preserving half of the supersymmetries and that asymptote to $\adsf{} \times S^5$ has been established \cite{Lin:2004nb}. The geometries are fully specified by a black and white coloring of a 2D plane, with the shapes of the black droplets specifying the topology and boundary conditions of the metric, mirroring the free-fermion description of \halfBPS{} states in  \Nfour{} SYM \cite{Berenstein:2004kk}.
    Recently, the heavy states dual to these geometries have been explicitly identified as coherent states:
    \begin{equation}
        \ket{\Lambda} = \int_{SU(N)} dU \exp[N \Tr(U X U^\dagger \Lambda)],
        \label{eq:half_coherent}
    \end{equation}
    where $X$ is one of the complex scalars of $\mathcal{N}=4$ SYM, and the eigenvalues of the matrix $\Lambda$ depend on the shape of the droplets \cite{Berenstein:2022srd}. This description has also been used to reproduce the three-point functions computed holographically \cite{Anempodistov:2025maj}.
    
    A particularly simple class of \halfBPS{} solutions is generated by elliptic droplet shapes. These geometries, that we shall review in section~\ref{sub:half_bps}, fit inside a five-dimensional consistent truncation of Type IIB supergravity. They have a simple CFT description, as an exponential over the lightest chiral primary:
    \begin{equation}
        \exp(\epsilon \Tr X^2),
    \end{equation}
    where $\epsilon$ is a deformation parameter, related to the eccentricity of the elliptical droplet. They can also be described as coherent states of the form \eqref{eq:half_coherent}, with a semi-circle distribution of eigenvalues for $\Lambda$. These geometries and their duals have played a major role in the recent computations of LLLL correlation functions of multi-trace operators \cite{Giusto:2024trt,Aprile:2024lwy,Aprile:2025hlt}.

    In comparison, the understanding of the holographic dictionary for heavy \quartBPS{} states is still lacking. The general gravity description is much more involved \cite{Donos:2006iy,Chen:2007du,Lunin:2008tf}, with the equations of motion reducing to non-linear Monge-Ampère-type systems. The solutions are specified by higher-dimensional droplets, and some attempts have been made to identify 1/4-BPS droplets with Young diagrams \cite{Kimura:2011df}. A simpler description of \quartBPS{} coherent heavy states, which parallels the 1/2-BPS description of \eqref{eq:half_coherent}, has recently been proposed in \cite{Anempodistov:2025maj}:
    \begin{equation}
        \ket{\Lambda_1,\Lambda_2} = \int_{SU(N)} dU \exp[N \Tr(U X U^\dagger \Lambda_1 + U Y U^\dagger \Lambda_2)] \,,
        \label{eq:quarter_coherent}
    \end{equation}
    where $\Lambda_1$ and $\Lambda_2$ are two diagonal matrices. Nonetheless their study is hindered by intractable matrix integrals, and no precise matching with bulk geometries exists.

    In this situation, a specific solvable instance of the \quartBPS{} dictionary could help clarify the duality. The goal of this paper is to study such an example. In doing so we provide the first matching between \quartBPS{} geometries and states in \Nfour{} SYM beyond the linearized level. We focus on a class of geometries, first studied in \cite{Chong:2004ce,Liu:2007rv}, that correspond in the language of \cite{Chen:2007du} to elliptical deformations of the 1/4-BPS droplet. These solutions fit inside the same consistent truncation as the elliptical \halfBPS{} geometries, which makes them amenable to computations, and readily provides substantial holographic data: the expectation values of all single-particle chiral primary operators of dimension higher than 2 vanish.
    
    The equations governing these geometries are not analytically solvable for finite deformations of the droplet. Nevertheless, we will able to extract the asymptotic behavior of all supergravity fields perturbatively. These encode for the three-point functions of dimension 2 chiral primaries, that we will compute following the dictionary of \cite{Skenderis:2006uy,Skenderis:2007yb}.

    Three-point functions involving \quartBPS{} operators are not generally protected by supersymmetry, except in special cases \cite{Bissi:2021hjk}. We propose a perturbative ansatz for the CFT dual to these geometries and show that all three-point functions involved to quadratic order are indeed protected. This data will then be used to fix the coefficients of the ansatz, leading to a precise determination of the CFT state to quadratic order, and making explicit the contribution of the lightest pure \quartBPS{} state \cite{DHoker:2003csh,Ryzhov:2001bp} to the quadratic piece of the heavy state. 

    The paper is organized as follows. In Section~\ref{sec:sym}, we review the spectrum of \halfBPS{} and \quartBPS{} operators of $\mathcal{N}=4$ SYM, highlighting the lightest \quartBPS{} state that emerges in our analysis. Section~\ref{sec:sugra} sets up the supergravity problem in terms of the consistent truncation. We review the supersymmetric analysis of these geometries and introduce a new ansatz that decouples the BPS equations and simplifies the perturbation theory. In Section~\ref{sec:cft}, we use the holographic dictionary to extract CFT expectation values and fix the coefficients of a perturbative ansatz for the heavy state, revealing the explicit contribution of the lightest \quartBPS{} operator. Section~\ref{sec:numerics} validates our results numerically, solving the full nonlinear equations and exploring the regime beyond perturbation theory. In Section~\ref{sec:conclusion}, we discuss our results and present future directions of research.

    \section{Spectrum of \Nfour{} SYM and holography}
    \label{sec:sym}

    In this section, we review the spectrum of \Nfour{} SYM with gauge group $SU(N)$. We also introduce the notations for the operators used in the rest of the paper.
    We are interested in the scalar sector of the theory. We recall that \Nfour{} SYM has 6 hermitian scalar fields $\Phi^i$, and it is useful to form the following complex combinations:
    \begin{equation}
        X = \frac{1}{\sqrt{2}} (\Phi^1 + i \Phi^2) \,,\quad \bar X = \frac{1}{\sqrt{2}} (\Phi^1 - i \Phi^2) \,,
    \end{equation} 
    and similarly for the pairs $Y, \bar Y$ and $Z, \bar Z$, constructed from $\Phi^3, \Phi^4$ and $\Phi^5, \Phi^6$, respectively.
    These fields transform in the adjoint of the gauge group $SU(N)$ and invariant operators can be formed by taking traces of products of these fields. They are normalized so that
    \begin{equation}
        \expval{\bar X^{i_1}_{\ j_1}(x) X^{i_2}_{\ j_2}(y)} = \frac{\delta^{i_1}_{\ j_2} \delta^{i_2}_{\ j_1} - \frac{1}{N} \delta^{i_1}_{\ j_1}\delta^{i_2}_{\ j_2}}{\abs{x-y}^2} \,.
    \end{equation}

    The single trace chiral primary operators of dimension $\Delta = J = k$ are defined as
    \begin{equation}
        \mathcal{T}_{k,I} = C^I_{i_1, \dots, i_k} \Tr(\Phi^{i_1} \dots \Phi^{i_k}) \,,
        \label{eq:single_trace_chirals}
    \end{equation}
    where $C^I_{i_1, \dots, i_k}$ is a totally symmetric rank-$k$ tensor of $SO(6)$ and $I$ is a generic label for the R-symmetry quantum numbers. These operators transform in the $[0,k,0]$ representation of $SO(6)$. Multi-trace chiral primaries can be formed as products of single-trace operators, and together they form the spectrum of \halfBPS{} operators.
    
    At leading order in $1/N$, the gravity dual of light single-trace operators have been identified with the linearized deformations of $\adsf{}\times S^5$ in \cite{Kim:1989abc}. However, strong arguments have been made that another basis of \halfBPS{} operators, made of single-particle operators, is more naturally dual to these fluctuations, and can account for $1/N$ corrections \cite{Aprile:2018efk,Aprile:2020uxk}. These single-particle operators are  \halfBPS{} operators that are defined to be orthogonal to all multi-trace operators. 
    In the context of this paper, they will allow us to rewrite the complicated constraints originating from the consistent truncation of section~\ref{sub:truncation} in a simple form.
    The lightest single-particle operators are
    \begin{equation}
        \Op_{2,I} = \mathcal{T}_{2,I} \qand \Op_{4,I} = \mathcal{T}_{4,I} -  \frac{2 N^2-3}{N (N^2+1)} \mathcal{T}_{2,I}^2  \,.
        \label{eq:chiral_primary}
    \end{equation}

    The construction of \quartBPS{} operators is more intricate. These operators belong to representations $[q,p,q]$ with $q>0$ and have protected conformal dimension $\Delta = 2q + p$. Systematic methods have been devised for their construction \cite{DHoker:2003csh,Ryzhov:2001bp}, and an orthogonal basis of all such operators has been constructed in terms of Young diagrams \cite{Lewis-Brown:2020nmg}. The lightest \quartBPS{} operator, transforming in the $[2,0,2]$ representation, plays a key role in this paper.
    In terms of elementary fields, it is given by\footnote{The expression for generic polarization can be found is section 4 of \cite{Bissi:2021hjk}.}
    \begin{equation}
        \Op_{1/4} = \Tr X^2 \Tr Y^2 - (\Tr XY)^2 - \frac2N (\Tr XXYY - \Tr XYXY))
        \label{eq:pure_quarter_state}
    \end{equation}
    This operator has protected dimension $\Delta = 4$ at one loop \cite{Ryzhov:2001bp}, and it is conjectured that it does not receive any corrections at higher loops  \cite{Pasukonis:2010rv}.

    In this paper, we are interested in the CFT dual of specific 1/2- and 1/4-BPS geometries asymptotic to \adsf{}, detailed in sections~\ref{sec:sugra}. Following the dictionary, these are heavy, coherent states that depend on the deformation parameters specifying the bulk geometries. In the perturbative regime, these states can be formally expanded as
    \begin{equation}
        \Op_H(\epsilon) = 1 + \sum_{n > 0} \epsilon^n \Op_n \,.
    \end{equation}
    As we will see, by making suitable ansätze for the operators $\Op_n$ in terms of products of the light operators described previously, it will be possible to use holography to determine the heavy states. In the \halfBPS{} case, this task has been achieved to all orders \cite{Aprile:2025hlt}, and the series has been resummed:
    \begin{equation}
        \Op_H^{(1/2)} = \exp(\epsilon \Tr X^2) \,.
    \end{equation}
    The 1/4-BPS case is more involved. Our task in this paper is to determine the operator appearing at quadratic order in the expansion, and to express it in terms of the lightest \quartBPS{} operator $\Op_{1/4}$.

    \section{Static AdS bubbles}
    \label{sec:sugra}

    \subsection{The five-dimensional truncation}
    \label{sub:truncation}

    We will be working with a consistent truncation of Type IIB supergravity on $S^5$ \cite{Cvetic:2000nc}. The theory has gauge group $SO(6)$ inherited from the symmetries of the 5-sphere. In addition to the 5D metric, it contains 20 hypermatter scalars, which transform in the $20'$ of $SO(6)$ and are represented by a symmetric tensor of unit determinant $T_{ij}$, where indices $i, j, \dots$ are the vector representation of $SO(6)$. It also contains 15 gauge fields, represented by an antisymmetric matrix of 1-forms $A_{ij}$.
    The Lagrangian takes a simple form:
	\begin{equation}
        \begin{aligned}
            \mathcal{L} &= eR - \frac e4 \Tr(T^{-1} (D_\mu T) T^{-1} (D^\mu T)) + \frac e8 \Tr(T^{-1} F_{\mu\nu} T^{-1} F^{\mu\nu}) - e V + \mathcal{L}_\text{Chern-Simons}
        \end{aligned}
	\end{equation}
    where $e= \sqrt{\det \abs{g}}$, $R$ is the Ricci scalar, and the scalar potential is given by
    \begin{equation}
        V = \Tr T^2 - \frac{1}{2} (\Tr T)^2 \,.
    \end{equation}
    The covariant derivative and field strength that appear in the kinetic terms for the scalars and gauge fields are
    \begin{equation}
        DT_{ij} = dT_{ij} + A^{ik} T_{kj} + A^{jk} T_{ik} \,,\quad F^{ij} = dA^{ij} + A^{ik} \wedge A^{kj} \,.
    \end{equation}
    The Chern-Simons interactions do not play a role in the geometries considered in this work. Their expression can be found in \cite{Cvetic:2000nc}.

    The solutions of this truncation can be uplifted to exact solutions of type IIB supergravity \cite{Cvetic:2000nc}.
    Parametrizing the 5-sphere by Cartesian coordinates $y^i$ ($i = 1, \dots, 6$) satisfying $y^i y^i = 1$, the ten-dimensional metric takes the form
    \begin{equation}
        ds_{10}^2 = \Delta^{1/2} ds_5^2 + \Delta^{-1/2} T_{ij}^{-1} Dy^i Dy^j \,,
        \label{eq:uplift_metric}
    \end{equation}
    where $ds_5^2$ is the asymptotically AdS Einstein metric of the 5D base space, $\Delta \equiv T_{ij} y^i y^j$ is a warp factor and $Dy^i = dy^i + A^{ij} y^j$ is the $SO(6)$-covariant derivative.
    Note that we work with dimensionless coordinates and have fixed $R_{\ads} = 1$.
    The five-form field strength is given by\footnote{Our conventions for the normalization of the five-form align with \cite{Skenderis:2007yb}, which differ from \cite{Cvetic:2000nc} by a factor of $4$.}
    \begin{align}
        &F_5 =  H_5 + \star H_5 \,, \label{eq:uplift_five_form}
        \\
        &\begin{aligned}
            H_5 = &-\frac{1}{4} (2 T_{ij} T_{jk} y^i y^k - \Delta \, T_{ii}) \varepsilon_5 + \frac{1}{4}  T_{ij}^{-1} y^k \star DT_{jk} \wedge Dy^i \\
            & - \frac{1}{8} T_{ik}^{-1} T_{jl}^{-1} \star F^{ij} \wedge Dy^k \wedge Dy^l \,,
        \end{aligned}
    \end{align}
    where $\varepsilon_5$ is the volume form of the 5D base space.

    For our specific application to \halfBPS{} and \quartBPS{} geometries, we further truncate the theory, by retaining only the diagonal subset of scalars $T_{ij}$:
    \begin{equation}
        T_{ij} = \diag(e^{-\mu_1 + \lambda_1}, e^{-\mu_1 - \lambda_1},e^{-\mu_2 + \lambda_2}, e^{-\mu_2 - \lambda_2},e^{-\mu_3 + \lambda_3}, e^{-\mu_3 - \lambda_3})
    \end{equation}
    where $\mu_1 + \mu_2 + \mu_3 = 0$. We also retain the three gauge fields on the maximal torus of $SO(6)$:
    \begin{equation}
        A^{12} = A_1 \,,\quad A^{34} = A_2 \,,\quad A^{56} = A_3 \,.
    \end{equation}
    This restricted ansatz preserves an $U(1)^3$ subgroup of the full $SO(6)$ symmetry and will prove sufficient to capture the essential physics of the geometries we wish to study. The remaining fields $\mu_i$, $\lambda_i$ and $A_i$, along with the five-dimensional metric, will form the basis of our analysis in the following sections, where we construct and analyze the supersymmetric solutions of interest.

    \subsection{Supersymmetric analysis}
    \label{sub:pope_analysis}

    A first construction of supersymmetric solutions in this 5D truncation was carried out in \cite{Chong:2004ce}, and the supersymmetries were fully analyzed a few years later in \cite{Liu:2007rv}. The most general static, smooth bubbling solutions are characterized by the following metric and field configurations:
    \begin{gather}
        ds_5^2 = -(H_1 H_2 H_3)^{-2/3} f dt^2 + (H_1 H_2 H_3)^{1/3} \qty(f^{-1} \frac{dx^2}{4x} + x \,d\Omega_3^2) \label{eq:susy_pope_1}
        \\
        A_i = - H_i^{-1} dt \,,\quad e^{-\mu_i} = (H_1 H_2H_3)^{1/3} H_i^{-1}  \,,\quad \cosh\lambda_i = (x H_i)' \label{eq:susy_pope_2}
    \end{gather}
    where 
    \begin{equation}
        f = 1 + x H_1 H_2 H_3 \,.
        \label{eq:f_pope}
    \end{equation}
    The functions $H_i$ depend only on the radial coordinate $x$ and must satisfy the equations of motion
    \begin{equation}
        f (x H_i)'' = -\frac{1}{2}\qty[(xH_i)'^2-1]\epsilon_{ijk} H_j H_k \,.
        \label{eq:eom_pope}
    \end{equation}

    These solutions preserve in general 1/8 of the supersymmetries of the vacuum. By setting some of the $H_i$ to $1$, the amount of preserved supersymmetries is increased.  In the single-charge case, $H_2 = H_3 = 1$, the equations can be solved analytically. These \halfBPS{} solutions are specific instances of LLM geometries \cite{Lin:2004nb}, with an elliptical droplet profile. Explicitly, one has
    \begin{equation}
        H_1 = \sqrt{1 + \frac{2(1+q_1)}{x} + \frac{1}{x^2}} - \frac{1}{x} \,,
        \label{eq:sol_half_pope}
    \end{equation}
    where $q_1$ is the R-charge that determines the strength of the elliptical deformation.
    
    The focus for the rest of the paper will be on \quartBPS{} solutions, which arise when $H_3 = 1$ while $H_1$ and $H_2$ are non-trivial functions of $x$. We will develop a suitable ansatz  for their study, construct the geometries both perturbatively and numerically, and ultimately identify their dual descriptions in the conformal field theory. 

    \subsection{A suitable ansatz}

    To facilitate both perturbative and numerical analysis of the \quartBPS{} solutions, we introduce a new ansatz for the five-dimensional fields that offers several advantages over the  formulation of Section \ref{sub:pope_analysis}. 
    Most notably, it allows for the decoupling of the BPS equations, provides a cleaner separation of corrections at different orders in perturbation theory, and offers greater flexibility for potential extensions to non-BPS or non-static solutions.

    The ansatz is constructed by first imposing the $SO(4) \times SO(2)$ symmetry characteristic of the \quartBPS{} solutions. In particular, the invariance under the $SO(2)$ subgroup leads to:
    \begin{equation}
        \lambda_3 = 0 \qand A^{56} = a_3\,dt \,,
    \end{equation} 
    where $a_3$ is an arbitrary constant that does not play a role in the dynamics, that we set to $a_3 = 1$.
    The remaining fields of the consistent truncation are parameterized as:
	\begin{align}
		ds_5^2 &= \omega_0^2 \qty[\frac{d\xi^2}{(1-\xi^2)^2} + \frac{\xi^2}{1-\xi^2} \frac{\omega_1^2}{\omega_1^2 + \xi^2 ( \omega_0^2 - \omega_1^2 )} d\Omega_3^2] - \frac{\omega_1^2}{1-\xi^2} dt^2 \,,
		\label{eq:metric_ansatz}
        \\
        A^{12} &= \phi_1 dt \,,\quad A^{34} = \phi_2 dt \,,
        \label{eq:gauge_ansatz}
        \\
        T_{ij} &= \diag(e^{-\mu_1 + \lambda_1}, e^{-\mu_1 - \lambda_1},e^{-\mu_2 + \lambda_2}, e^{-\mu_2 - \lambda_2},e^{\mu_1 + \mu_2}, e^{\mu_1 + \mu_2})
        \label{eq:scalar_ansatz}
	\end{align}
    where $\xi$ is a compact radial coordinate ranging from $0$ to $1$, with $\xi = 1$ corresponding to the boundary of \ads{}. The fields $\lambda_i, \phi_i, (i=1,2)$ and $\omega_0,\omega_1$ are all functions of the radius $\xi$.
    The particular factor multiplying the line element of the 3-sphere fixes the reparametrization invariance $\xi \mapsto \xi'(\xi)$, and its form is inspired by similar ansätze involved in the construction of \quartBPS{} and non-BPS solutions in $\ads{}_3$ \cite{Ganchev:2023sth}. This choice is advantageous for both perturbation theory and numerical analysis: as we will see, at each order in perturbation theory, the fields are polynomials of $\xi$.
    In this language, the vacuum solution is given by:
    \begin{equation}
        \begin{gathered}
            \omega_0 = \omega_1 = 1 \,,\\
            \phi_i = 1 \,,\quad \lambda_i = \mu_i = 0 \,,\quad i = 1,2 \,.
        \end{gathered}
    \end{equation}

    \subsection{The BPS equations}
	\label{sub:BPSeqs}

    One way to determine the BPS equations of this ansatz is to match the metric \eqref{eq:metric_ansatz}, the gauge fields \eqref{eq:gauge_ansatz} and the scalars \eqref{eq:scalar_ansatz} with the expressions of \eqref{eq:susy_pope_1} and \eqref{eq:susy_pope_2}. We first straightforwardly identify the gauge fields and scalar fields:
    \begin{gather}
        \phi_i = H_i^{-1}\,, \quad i = 1, 2 \,, \label{eq:phi_to_pope}\\
        e^{-\mu_1} = \frac{\phi_1^{2/3}}{\phi_2^{1/3}} \,,\quad e^{-\mu_2} = \frac{\phi_2^{2/3}}{\phi_1^{1/3}} \,.
        \label{eq:mu_BPS}
    \end{gather}
    
    Turning to the identification of the metrics, we find the expression of the radial coordinate $x$ in terms of $\xi$
    \begin{equation}
        x = \frac{\xi^2}{(1-\xi^2)} \frac{\omega_0^2 \omega_1^4}{\qty((1-\xi^2) \omega_1^2 + \xi^2 \omega_0^2)^{3/2}} \,.
    \end{equation}
    The fact that this matching is solution-dependent explains why certain quantities are more conveniently computed in the new ansatz.
    From \eqref{eq:f_pope} and the $dt^2$ and $dx^2$ terms in \eqref{eq:susy_pope_1}, we obtain two equivalent expressions for $f$,
    \begin{equation}
        f = 1 + \frac{\xi^2 \omega_0^2}{(1-\xi^2)\omega_1^2} = 1 + \frac{\xi^2}{(1-\xi^2)} \frac{\omega_0^2 \omega_1^4}{\phi_1 \phi_2 \qty((1-\xi^2) \omega_1^2 + \xi^2 \omega_0^2)^{3/2}}  \,.
    \end{equation}
    They can be combined with the scalar field relations \eqref{eq:mu_BPS} to obtain another algebraic BPS constraint, which can be used to fix $\omega_0$:
    \begin{equation}
        \xi^2 \omega_0^2 + \omega_1^2 (1-\xi^2) - e^{2(\mu_1+\mu_2)} \omega_1^4 = 0
        \label{eq:omega0_bps}
    \end{equation}
    Then, the last equations of \eqref{eq:susy_pope_2} reduce to first-order BPS equations:
    \begin{equation}
        \cosh\lambda_i = \phi_i^{-1} + G^{-1} \phi_i^{-2} \xi \partial_\xi \phi_i
        \label{eq:phi_BPS}
    \end{equation}
    where we have defined
    \begin{equation}
        G \equiv e^{2(\mu_1+\mu_2)} \frac{\omega_1^2}{1-\xi^2} \,.
        \label{eq:omega1_bps}
    \end{equation}
    The equations of motion \eqref{eq:eom_pope} are equivalent to the following first-order BPS equations:
    \begin{equation}
        \frac{\xi \partial_\xi \lambda_i}{\sinh \lambda_i} = -2 \phi_i (G-1) \,.
        \label{eq:lambda_bps}
    \end{equation}

    Interestingly, substituting these BPS constraints into the 5D equations of motion, one can derive an integral of the motion:
    \begin{equation}
        \partial_\xi C = 0 \,,\qquad C \equiv \frac1{\xi^2} G (G-1) \sinh\lambda_1 \sinh\lambda_2\, \phi_1 \phi_2  \,.
        \label{eq:integral_BPS}
    \end{equation}

    To decouple the equations, we introduce new fields
    \begin{equation}
		\sigma_{\pm} = \log\tanh(\frac12 \lambda_1) \pm \log\tanh(\frac12 \lambda_2) \,.
        \label{eq:eom_sigma_pm}
	\end{equation}
    and using \eqref{eq:phi_BPS} and \eqref{eq:lambda_bps}, one finds that they satisfy the decoupled equations\footnote{These equations are identical to the ones governing the double superstratum breathing mode of \cite{Ganchev:2023sth} (eq. (2.55),(2.56)).}
    \begin{equation}
		\xi \partial_\xi (\xi\partial_\xi \sigma_{\pm}) - 4 C \xi^2 \sinh \sigma_{\pm} = 0 \,,
        \label{eq:decoupled_bps}
	\end{equation}
    All other fields can be expressed in terms of $\sigma_{\pm}$ and their $\xi$-derivatives:
    \begin{equation}
        \begin{aligned}
            \lambda_1 &= 2 \tanh ^{-1}\qty(e^{\frac{1}{2} (\sigma_+ + \sigma_-)}) \,,\quad 
            \lambda_2 = 2 \tanh ^{-1}\qty(e^{\frac{1}{2} (\sigma_+ - \sigma_-)}) \,,
            \\
            \phi_1 + \phi_2 &= \qty(\frac{1}{2}  - 4C\,\frac{\cosh\sigma_+ - \cosh\sigma-}{\sigma_+'^2 - \sigma_-'^2})\xi \sigma_+' \,,
            \\
            \phi_1 - \phi_2 &= \qty(\frac{1}{2}  - 4C\,\frac{\cosh\sigma_+ - \cosh\sigma-}{\sigma_+'^2 - \sigma_-'^2})\xi \sigma_-' \,,
            \\
            % e^{\mu_1 - \mu_2} &= 1 - \frac{2 \sigma_-'}{\sigma_+' + \sigma_-'} \,,
            % \\
            % e^{-3(\mu_1 + \mu_2)} &= - \frac{\xi^2 \qty(8 C (\cosh \sigma_+ - \cosh\sigma_-) - \sigma_+'^2 + \sigma_-'^2)^2}{16 (\sigma_+'^2 - \sigma_-'^2)} \,,
            % \\
            e^{-3\mu_1} &= \frac{1}{4} \xi \frac{(\sigma_+' + \sigma_-')^2}{\sigma_+' - \sigma_-'} - 2C \xi \, \frac{(\cosh\sigma_+ - \cosh\sigma_-)(\sigma_+' + \sigma_-')}{(\sigma_+' - \sigma_-')^2}
            \\
            e^{-3\mu_2} &= \frac{1}{4} \xi \frac{(\sigma_+' - \sigma_-')^2}{\sigma_+' + \sigma_-'} - 2C \xi \, \frac{(\cosh\sigma_+ - \cosh\sigma_-)(\sigma_+' - \sigma_-')}{(\sigma_+' + \sigma_-')^2}
            \\
            G &= \qty(1 - \frac{\sigma_+'^2 - \sigma_-'^2}{8C(\cosh\sigma_+ - \cosh\sigma_-)})^{-1}
        \end{aligned}
        \label{eq:fields_in_sigma}
    \end{equation}
    and $\omega_0$, $\omega_1$ can be reconstructed from \eqref{eq:omega0_bps} and \eqref{eq:omega1_bps}.

    The \quartBPS{} geometries are thus completely determined by the solutions of \eqref{eq:decoupled_bps}, subject to the boundary conditions:
    \begin{equation}
        \sigma_{\pm} =_{\xi \to 0}  \bigO(1) \,,\quad \sigma_- =_{\xi \to 1} \bigO(1) \qand \sigma_+ =_{\xi \to 1} \log(\frac{C}{4}\qty(1-\xi^2)^2) + \bigO(1-\xi^2) 
        \label{eq:sigma_bdy}
    \end{equation}
    where the last condition comes from \eqref{eq:integral_BPS}. Unfortunately, there is no closed-form expression for the solutions of the differential equation \eqref{eq:decoupled_bps}. These fields are also poorly adapted to the perturbation theory, as $\sigma_+$ diverges close to the vacuum solution ($C = 0$), and in the next section we will prefer individual expansions of all the fields.

    \subsection{Review of half-BPS geometries}
    \label{sub:half_bps}

    The \halfBPS{} solutions have an additional $SO(2)$ invariance that imposes $\lambda_2 = 0$, $\phi_2 = 1$ and, from \eqref{eq:mu_BPS}, $\mu_2 = -\mu_1 / 2$. In that case, the constant $C$ is trivially zero, but one can find an additional integral of the motion,
    \begin{equation}
        \partial_\xi\hat C = 0 \,,\qquad \hat C \equiv G \sinh \lambda_1 \,,
    \end{equation}
    which can be used to substitute $G$ in \eqref{eq:phi_BPS} and \eqref{eq:lambda_bps}. One can then integrate these equations, and after some manipulations one obtains that
	\begin{equation}
		\begin{aligned}
		    e^{\lambda_1} &= e^{2\beta_1}\, \frac{\cosh\beta_1 - \xi^2 \sinh\beta_1}{\cosh\beta_1 + \xi^2 \sinh\beta_1} \,, \quad
			\phi_1^{-1} = e^{3\mu_1} = \xi^2 + (1-\xi^2) \cosh2\beta_1 \,,
			\\
			\omega_1^2 &= e^{-2\mu_1} \,\frac{\cosh^2\beta_1 - \xi^4 \sinh^2\beta_1}{ \cosh^2\beta_1 - \xi^2 \sinh^2\beta_1} \,,
			\quad
			\omega_0^2 = e^{\mu_1} \,\frac{\cosh^2\beta_1 - \xi^4 \sinh^2\beta_1}{\qty(\cosh^2\beta_1 - \xi^2 \sinh^2\beta_1)^2}\,.
		\end{aligned}
		\label{eq:half_bps_solution}
	\end{equation}
    where $\beta_1$ is defined by $\hat C = \sinh 2\beta_1$. It is easy to see that this solution is identical to the LLM elliptical solution \eqref{eq:sol_half_pope}, with the R-charge given by
    \begin{equation}
        q_1 = 2 \sinh^2 \beta_1 \,.
    \end{equation}

    \subsection{Perturbative expansion of the quarter-BPS geometries}
    \label{sub:perturbations}
    Given the coupled and nonlinear nature of both the equations of motion and BPS equations \eqref{eq:eom_pope}, exact analytical solutions remain elusive. In the form \eqref{eq:decoupled_bps}, the BPS equations can be decoupled, at the price of increasing the order of the differential equations, yet even these simplified equations cannot be solved analytically. We therefore adopt a perturbative approach to construct the quarter-BPS geometries.
    We start with the global \adsf{} solution, and assume that the fields are expanded in powers of a bookkeeping parameter $\epsilon$:
    \begin{equation}
        \begin{aligned}
            \lambda_i &= \sum_{n\geq 0} \epsilon^{2n+1} \,\delta^{(2n+1)}\lambda_i  \,,\quad \mu_i = \sum_{n>0} \epsilon^{2n} \,\delta^{(2n)}\mu_i \,,\quad \phi_i = 1 + \sum_{n>0} \epsilon^{2n} \,\delta^{(2n)}\phi_i  \,,\quad i=1,2
            \\
            \omega_0 &= 1 + \sum_{n>0} \epsilon^{2n} \,\delta^{(2n)}\omega_0 \,,\quad \omega_1 = 1 + \sum_{n>0} \epsilon^{2n} \,\delta^{(2n)}\omega_1 \,.
        \end{aligned}
    \end{equation}

    At linear order, only the fields $\lambda_1$ and $\lambda_2$ are excited, their linearized equations of motion are
    \begin{equation}
        \xi  \left(1-\xi ^2\right)^2 \delta^{(1)}\lambda _i'' + \left(1-\xi ^2\right) \left(3-\xi ^2\right) \delta^{(1)} \lambda_i' - 4 \xi  \left(2-\xi ^2\right) \delta^{(1)}  \lambda_i = 0 \,.
    \end{equation}
    with normalizable excitations given by
    \begin{equation}
    	\label{eq:lambda12Norm}
        \delta^{(1)}\lambda_1 = 2 \beta_1 (1-\xi^2) \,,\quad \delta^{(1)}\lambda_2 = 2 \beta_2 (1- \xi^2) \,,
    \end{equation}
    where $\beta_1$ and $\beta_2$ are the amplitudes of the deformations, and the factors of $2$ are introduced for notational convenience. Note that fixing $\beta_2 = 0$ yields the \halfBPS{} solution \eqref{eq:half_bps_solution}.

    Higher-order corrections can be determined iteratively by solving the linearized equations sourced by lower-order terms. At each order, we find a unique smooth solution respecting the boundary conditions, with the fields at each order expressed as polynomials of the radius $\xi$. Importantly, we encounter no obstructions that would prevent extending this procedure to arbitrarily high orders in $\epsilon$. For comparisons with the numerical results of Section~\ref{sec:numerics}, we computed all fields up order $12$.
    
    We present here the results up to fourth order.  For conciseness, we do not write down the expansion of the fields $\lambda_2$, $\mu_2$ and $\phi_2$, as those can be directly obtained from $\lambda_1$, $\mu_1$ and $\phi_1$ by exchanging $\beta_1$ and $\beta_2$, respectively. At second order, we find:
    \begin{equation}
        \begin{gathered}
            \delta^{(2)} \mu_1 = \frac{2}{3} (2\beta_1^2 - \beta_2^2) (1 - \xi^2) \,,\quad
            \delta^{(2)} \phi_1 = -2 \beta_1 ^2 (1-\xi^2) \,,\\
            \delta^{(2)} \omega_0 = -\frac{1}{6}(\beta_1^2+\beta _2^2) \left(1-\xi ^2\right) \left(1-3 \xi ^2\right) \,,\\
            \delta^{(2)} \omega_1 = -\frac{1}{6}(\beta_1^2+\beta _2^2) \left(1-\xi ^2\right) \left(4-3 \xi ^2\right) \,.
        \end{gathered}
    \end{equation}
    At third order (and all odd orders), only the scalar fields $\lambda_i$ receive corrections:
    \begin{equation}
        \delta^{(3)} \lambda_1 = \frac{2}{3} \beta _1  \left(\beta _1^2 \left(1+\xi ^2\right)-3 \beta _2^2\right) \xi ^2 \left(1-\xi ^2\right) \,.
    \end{equation}
    At fourth order, the corrections become more involved:
    \begin{equation}
        \begin{aligned}
            \delta^{(4)} \mu_1 &= \frac{2}{9} \left(1-\xi ^2\right) \left(\beta _1^4 \left(6 \xi ^2 - 4\right)-\beta _1^2 \beta _2^2 \xi ^2 \left(\xi ^2+3\right)-\beta _2^4 \left(3 \xi ^2-2\right)\right) \,,
            \\
            \delta^{(4)} \phi_1 &= \frac{2}{3} \left(1-\xi ^2\right) \left(\beta _1^4 \left(-6 \xi ^2 + 5\right)+\beta _1^2 \beta _2^2 \xi ^2 \left(\xi ^2+3\right)\right) \,,
            \\
            \delta^{(4)} \omega_0 &= \frac{1}{72} \left(1-\xi ^2\right) \big[\beta _1^4 \left(9 \xi ^6+33 \xi ^4-25 \xi ^2-9\right)-2 \beta _1^2 \beta _2^2 \left(9 \xi ^6+61 \xi ^4-29 \xi ^2-9\right)\\
            &\quad +\beta _2^4 \left(9 \xi ^6+33 \xi ^4-25 \xi ^2-9\right)\big] \,,
            \\
            \delta^{(4)} \omega_1 &= \frac{1}{72} \left(1-\xi ^2\right) \big[\beta _1^4 \left(9 \xi ^6+51 \xi ^4-112 \xi ^2+48\right)-2 \beta _1^2 \beta _2^2 \left(9 \xi ^6+25 \xi ^4-34 \xi ^2-16\right)\\
            &\quad +\beta _2^4 \left(9 \xi ^6+51 \xi ^4-112 \xi ^2+48\right)\big] \,.
        \end{aligned}
    \end{equation}
    These perturbative results provide the necessary input for the holographic analysis in the following section.

    \section{Identifying the CFT dual}
    \label{sec:cft}

    The \quartBPS{} geometries are dual to coherent heavy states involving a large number of traces. Based on the symmetries of the problem, we expect its constituents to be products of single-trace operators involving the scalars $X$ and $Y$. 
    To construct the dual CFT state, we employ a perturbative approach. We build an ansatz that includes all multi-trace operators allowed by the symmetries, with undetermined coefficients. These coefficients will then be fixed by matching protected quantities between the free theory and the supergravity regime.

    A particularly useful set of protected quantities are the expectation values of light chiral primaries, which can be computed in the supergravity regime following the holographic dictionary described in \cite{Skenderis:2006uy,Skenderis:2007yb}. In the CFT, these translate to three-point correlation functions involving one light \halfBPS{} state and two heavy \quartBPS{} states. 

    \subsection{Holographic expectation values}
    \label{sub:holography}

    From the behavior of the fields at the asymptotic boundary, one can extract the charges of the dual heavy state, as well as the expectation values of some chiral primaries. We shall review and apply the method of Kaluza-Klein holography \cite{Skenderis:2006uy,Skenderis:2007yb}. Our approach is solution-independent, yielding results applicable to any geometry satisfying the ansatz \eqref{eq:metric_ansatz}-\eqref{eq:scalar_ansatz}. We then specialize to \quartBPS{} geometries towards the end of the section.

    The first step of the holographic computation is to rewrite the metric \eqref{eq:metric_ansatz} in Fefferman-Graham coordinates:
    \begin{equation}
        ds_5^2 = \frac{dz^2}{z^2} + \frac{1}{z^2} \qty(-dt^2 + d\Omega_3^2) + \qty(g_{\mu\nu}^{(2)} + z^2 g_{\mu\nu}^{(4)} + \dots) dx^\mu dx^\nu \,,
    \end{equation}
    where $\mu, \nu$ are coordinates on the four-dimensional boundary, which is situated at $z=0$. This form of the metric  is an expansion close to the boundary, and the ``$\dots$'' includes terms with higher powers of $z$. It will be useful to introduce notations for the asymptotic behavior of the fields near the boundary, in the original radial coordinate $\xi$:
    \begin{equation}
        \mathcal{F}(\xi) \sim 
        %\sum_{n\geq 0} (1 - \xi^2)^n \mathcal{F}^{(n)}
        \mathcal{F}^{(0)} + (1 - \xi^2) \mathcal{F}^{(1)} + (1 - \xi^2)^2 \mathcal{F}^{(2)} + \dots
        \qq{as} \xi \sim 1 \,,
        \label{eq:bdy_expansion}
    \end{equation}
    where $\mathcal{F}$ represents any field in the ansatz $\{\lambda_i, \mu_i, \phi_i, \omega_0, \omega_1\}$. The boundary conditions on these fields are simply
    \begin{equation}
        \begin{gathered}
            \omega_0^{(0)} = \omega_1^{(0)} = 1\,,\quad \lambda_i^{(0)} = \mu_i^{(0)} = 0 \,,\quad \phi_i^{(0)} = 1 \,,\quad (i = 1,2)
        \end{gathered}
    \end{equation}
    One can then determine the relation between the bulk radius $\xi$ and the Fefferman-Graham coordinate $z$:
    \begin{equation}
        \xi = 1 - \frac{1}{2} z^2 + \frac{1}{8} \qty(1 + 4 \omega_0^{(1)}) z^4 + \bigO(z^6) \,.
    \end{equation}
    From this, we deduce that the first corrections to the boundary metric are given by
    \begin{equation}
        g_{\mu\nu}^{(2)} dx^\mu dx^\nu = -\frac{1}{2} (dt^2 + d\Omega_3^2) + \qty(\omega_0^{(1)} + 2 \omega_1^{(1)}) (-dt^2 + d\Omega_3^2)\,.
        \label{eq:FGorder2}
    \end{equation}
    In practice the second term vanishes, since $\omega_0^{(1)} = -2 \omega_1^{(1)}$ in all known solutions. Note however that this is not a direct consequence of the equations of motion. In what follows, we will also require the next order corrections:
    \begin{equation}
        \begin{aligned}
            g_{\mu\nu}^{(4)} &dx^\mu dx^\nu = \qty(\omega_0^{(1)} - 2 \omega_1^{(1)}) (dt^2 + d\Omega_3^2)  
            \\
            &+ \frac{1}{16} \qty(1 + 12 \omega_0^{(1)} - 8 (\omega_0^{(1)})^2 - 32 \omega_1^{(1)} + 16 (\omega_1^{(1)})^2 + 8 \omega_0^{(2)} + 32 \omega_1^{(2)} ) (-dt^2 + d\Omega_3^2) \,.
        \end{aligned}
        \label{eq:FGorder4}
    \end{equation}

    The next step of the holographic computation is to uplift the solutions to ten dimensions using the formulae \eqref{eq:uplift_metric}, \eqref{eq:uplift_five_form}.
    %, and to decompose the fluctuations of the 10D metric and five form in a basis of spherical harmonics.
    We denote by $h_{MN}$ and $f_{M_1\dots M_5}$ the fluctuations of respectively the metric and five-form around the vacuum solution:
    \begin{equation}
        \begin{aligned}
            ds_{10}^2 &= ds_{AdS_5}^2 + ds_{S^5}^2 + h_{MN} dx^M dx^N \,,\\
            F_5 &= \vol_{AdS_5} + \vol_{S^5} + \frac{1}{5!} f_{M_1 M_2 \dots M_5} dx^{M_1} \wedge \dots \wedge dx^{M_5} \,,
        \end{aligned}
    \end{equation}
    where we used $M,N,\dots$ for ten-dimensional indices, and from now on we also use $\mu, \nu, \dots$ for \ads{} indices and $a,b,\dots$ for $S^5$ indices. 
    We decompose these fluctuations in a basis of spherical harmonics. For our purposes, it is enough to consider the following fields and their KK decomposition\footnote{For a complete decomposition of all fluctuations, see \cite{Skenderis:2006uy}.}:
    \begin{align}
        h^a_a &= \sum \pi_{k, I_1} Y^{k, I_1} \,,
        \\
        h_{(ab)} &= \sum \Big(
            \hat{\phi}^{k, I_{14}}_{(t)}\, Y^{k, I_{14}}_{(ab)}
            + \phi^{k, I_5}_{(v)}\, D_{(a} Y^{k, I_5}_{b)}
            + \phi^{k, I_1}_{(s)}\, D_{(a} D_{b)} Y^{k, I_1}
            \Big) \,,
        \\
        h_{\mu a} &= \sum \left( B^{k,I_5}_{(v)\mu}\, Y^{k,I_5}_{a}
        + B^{k,I_1}_{(s)\mu}\, D_a Y^{k,I_1} \right) \,,
        \\
        f_{abcde} &= \sum b^{k, I_1}_{(s)}\, \Lambda_{k}\, \epsilon_{abcde}\, Y^{k, I_1} \,,
        \\
        f_{abcd\mu} &= \sum \epsilon_{abcd}{}^{e}\left(
        D_\mu b^{k,I_1}_{(s)}\, D_{e} Y^{k,I_1}
        + (\Lambda^{(v)}_{k} - 4)\, b^{k,I_5}_{(v)\mu}\, Y^{k,I_5}_{e}
    \right)
    \end{align}
    where all the dependence on the  5-sphere angles is encoded in the spherical harmonics $Y^{k, I_1}$, $Y^{k, I_5}$, $Y^{k, I_{14}}$. The parentheses surrounding indices, as in $h_{(ab)}$, denote the symmetric traceless combination. In Appendix~\ref{app:spherical_harmonics} we give more details about the harmonics, and explain how to extract the individual modes.

    The individual modes appearing in these formulae are not gauge-invariant. One way to carry out the computations is to impose the De Donder gauge fixing condition, which drastically simplifies the holographic dictionary. However, the metric \eqref{eq:metric_ansatz} is not in this gauge, and working out the change of variable is not trivial. We instead follow the procedure laid out in \cite{Skenderis:2006uy}, where the authors constructed combinations of modes that are gauge independent at linear order in the fluctuations. These are sufficient to compute the holographic expectation values of dimension-2 chiral primaries. The combinations we need are given by
    \begin{equation}
        \begin{gathered}
            \hat\pi_{k,I} = \pi_{k,I} - \Lambda_{k} \phi^{k,I}_{(s)} \,,\quad 
            \hat b_{(s)}^{k,I} = b_{(s)}^{k,I} - \frac{1}{2} \phi^{k,I}_{(s)} \,,
            \\
            \hat B_{(v)\mu}^{k,I_5} = B_{(v)\mu}^{k,I_5} - \frac{1}{2} D_\mu \phi_{(v)}^{k,I_5} \,,\quad
            \hat b^{k,I_5}_{(v)\mu} = b^{k,I_5}_{(v)\mu} - \frac{1}{2(\Lambda^{k,I_5} - 4)} D_\mu \phi_{(v)}^{k,I_5} \,.
        \end{gathered}
    \end{equation}
    From these, one can construct the set of scalar fields that diagonalize the five-dimensional equations of motion:
    \begin{equation}
        s_{k,I} = \frac{1}{20(k+2)} \qty(\hat\pi_{k,I} - 10(k+4) \hat b_{(s)}^{k,I}) \,,
    \end{equation}
    as well as the vector combination
    \begin{equation}
        a_\mu^{k,I_5} =  \hat B_{(v)\mu}^{k,I_5} - 4(k+3) \,.
    \end{equation}

    The holographic dictionary for the chiral primaries at $k=2$ then relates these fields to CFT expectation values\footnote{Note that the normalization of $\Op_{2,I}$ is different from the choice in \cite{Skenderis:2006uy}, hence the different prefactor. We have also corrected $N^2 \to N^2 - 1$ to account for the gauge group being $SU(N)$. This does not change the leading large-$N$ result, but as we will see in the next section, the equality is valid at finite $N$.}
    \begin{equation}
        \expval{\Op_{2,I}} = \frac{4}{3} (N^2 - 1) [s_{2,I}]_{2} \,,
    \end{equation}
    where the chiral primary operator $\Op_{2,I}$ was defined in \eqref{eq:chiral_primary}, and $[s_{2,I}]_{2}$ is the term of order $z^2$ in the Fefferman-Graham expansion of $s_{2,I}$.

    There are 10 independent dimension-2 $SO(2)$-invariant chiral primaries. Using the dictionary in Appendix~\ref{app:spherical_harmonics} we compute their expectation values at $t=0$, and express them in terms of the expansion of the fields at the boundary \eqref{eq:bdy_expansion}:
    \begin{equation}
        \begin{aligned}
            \expval{\Tr X^2} = \expval{\Tr \bar X^2} &= -\frac{1}{2} (N^2 - 1) \lambda_1^{(1)} \,,
            \\
            \expval{\Tr Y^2} = \expval{\Tr \bar Y^2} &= -\frac{1}{2} (N^2 - 1) \lambda_2^{(1)} \,,
            \\
            \expval{\Tr( X \bar X + Y \bar Y - 2 Z \bar Z)} &=  \frac{3}{2} (N^2 - 1) (\mu_1^{(1)} + \mu_2^{(1)}) \,,
            \\
            \expval{\Tr(X \bar X - Y \bar Y)} &= - \frac{1}{2} (N^2 - 1) (\mu_1^{(1)} - \mu_2^{(1)}) \,,
            \\
            \expval{\Tr XY} = \expval{\Tr \bar X \bar Y} &= \expval{\Tr \bar X Y} = \expval{\Tr X \bar Y} = 0\,.
        \end{aligned}
        \label{eq:vevs}
    \end{equation}
    The integrated R-charges of the solution can also be computed holographically, they are given by
    \begin{equation}
        \begin{aligned}
            \expval{J_1} &= - \frac{\sqrt{2}}{12} (N^2 - 1) [a_t^{1,\underline{1}}]_2 = -\frac{1}{2} (N^2 - 1) \phi_1^{(1)} \,,\\
            \expval{J_2} &= - \frac{\sqrt{2}}{12} (N^2 - 1) [a_t^{1,\underline{2}}]_2 = -\frac{1}{2} (N^2 - 1) \phi_2^{(1)} \,. 
            \label{eq:charges}
        \end{aligned}
    \end{equation}
    Finally the energy of the geometry is accessed through the holographic stress-tensor. Assuming that the second term in \eqref{eq:FGorder2} vanishes,  as it does for all geometries considered here, the result is
	\begin{equation}
        E = (N^2 - 1) \qty(g_{tt}^{(4)} - \frac{1}{4}g_{tt}^{(0)} g^{(4)}_{\ \,\lambda}{}^{\lambda}) = \frac32 (N^2 - 1) \qty(\omega_0^{(1)} - 2 \omega_1^{(1)}) \,,\\
        \label{eq:energy}
	\end{equation}
    where in the second equality we have used \eqref{eq:FGorder4}.

    Note that these results are valid for any solution that fit in the ansatz \eqref{eq:metric_ansatz}, \eqref{eq:gauge_ansatz}, \eqref{eq:scalar_ansatz}, and depend only on the expansion of the fields at first order at the boundary. In Appendix~\ref{app:vevs}, we derive all of these quantities in terms of the amplitude of the breathing mode perturbations, analytically for the \halfBPS{} geometries, and using the perturbative expansion for \quartBPS{} geometries. As consistency checks, we verified that $E = J_1 + J_2$ holds for all these geometries, and that the \halfBPS{} results are consistent with the identification of the heavy state in \cite{Aprile:2025hlt}.

    Regarding expectation values of higher-dimension operators, the consistent truncation projects out all the scalar fields but the ones in the $[0,2,0]$ representation. We therefore expect
    \begin{equation}
        \expval{\Op_p} = 0 \,, \quad p > 2 \,,
    \end{equation}
    where $\Op_p$ are single-particle operators of dimension $p$, as defined in section~\ref{sec:sym}. We verify this property on one example at dimension 4. Consider the  $SO(4)$-invariant, R-symmetry highest-weight dimension-4 chiral primary:
    \begin{equation}
        \Op_{4(4)} = \Tr X^4 - \frac{2 N^2-3}{N (N^2+1)} (\Tr X^2)^2 \,.
        \label{eq:highest_sp4}
    \end{equation}
    Computing its expectation value requires the gauge-invariant fluctuations of the fields at quadratic order, whose expressions are rather involved (see (3.19), (3.20) of \cite{Skenderis:2006uy}). We do not compute the full expression, but rather argue that the expectation value vanishes by a simple argument. Indeed, following the dictionary, $\expval{\Op_{4(4)}}$ is a sum of rank-4 modes of the fluctuations at order $z^4$, and of quadratics in the rank-2 modes at order $z^2$. Computing all these terms independently, we find that all of them are proportional to $(\lambda_1^{(1)})^2$, thus $\expval{\Op_{4(4)}} = \alpha (\lambda_1^{(1)})^2$, where $\alpha$ is a solution-independent coefficient. This formula holds for any geometry that fits the ansatz, and in particular for \halfBPS{} geometries, for which one indeed finds $\expval{\Op_{4(4)}} = 0$ \cite{Turton:2025svk}. Since $\lambda_1^{(1)}$ is non-zero in these geometries, we conclude that $\alpha = 0$, and thus 
    \begin{equation}
        \expval{\Op_{4(4)}} = 0
    \end{equation}
    for all geometries of the consistent truncation.

    \subsection{Identification of the half-BPS state}
    \label{sub:review_half_state}

    Before jumping to the identification the state dual to the quarter-BPS states, it is instructive to review the procedure in the simpler case of half-BPS geometries of section~\ref{sub:half_bps}, as presented in \cite{Turton:2025svk}. These geometries are dual to heavy states, whose dimension are proportional to $N^2$, and are made of a coherent sum of of multi-trace operators. Supersymmetry plays a crucial role in reducing the number of operators that can enter this sum: the spectrum of half-BPS operators is well understood.
    Thus the state is made of sums and products of single-trace operators of the form
    \begin{equation}
        \Tr X^k \,, k \geq 2 \,.
    \end{equation}
    Note that the polarization of these operators is fixed by the choice of parametrization of the 5-sphere in the bulk. 

    We proceed perturbatively in the parameter $\beta_1$. At linear order, one may use the textbook dictionary constructed in \cite{Kim:1989abc} to find:
    \begin{equation}
        \ket{\Psi} \sim 1 - \frac{1}{2} \beta_1 \Tr X^2 + \dots
    \end{equation}
    We use the symbol $\sim$ to indicate the equality up to an unspecified normalization constant. At higher orders, we can make the ansatz:
    \begin{equation}
        \begin{aligned}
        \ket{\Psi} ~\sim~ 1 & - \frac{1}{2} \beta_1 \Tr X^2 \\ 
        & + \beta_1^2 \qty[c_1 \Tr X^2 + c_2 (\Tr X^2)^2 + c_3 \Tr X^4] \\
        & + \beta_1^3 \qty[c_4 \Tr X^2 + \dots] + \mathcal{O}(\beta_1^4) \,.
        \end{aligned}
    \end{equation}
    where $c_1, c_2,  \dots$ are constants to be determined. Note that for technical reasons that will soon become clear, it is necessary to include $c_4$ in the ansatz to determine the state at quadratic order.

    The constants $c_j$ are fixed by matching the holographic expectation values computed in the bulk \eqref{eq:vevs}, with the free theory expectation values based oin the background given by the ansatz. Because of supersymmetry, these quantities are protected and do not depend on the coupling. In other words, the following equations must hold:
    \begin{equation}
        \frac{\expval{\Op_{I}}{\Psi}}{\braket{\Psi}} = \expval{\Op_{I}}_{\text{sugra}} \,,\quad
        \frac{\expval{J_1}{\Psi}}{\braket{\Psi}}  = \expval{J_1}_{\text{sugra}}% \,,\quad \frac{\expval{J_2}{\Psi}}{\braket{\Psi}}  = \expval{J_2}_{\text{gravity}}\,,
        \label{eq:match_vevs}
    \end{equation}
    where the left-hand side is computed with the ansatz in the free theory.  The free theory expectation values are readily computable via Wick contractions, see Appendix~\ref{app:correlators} for a list of three-point functions. The operator $\Op_{I}$ can be any single-trace chiral primary, but for our purposes we only need those of dimension $2$ and $4$. Although these equations are generically quadratic in the parameters of the ansatz, a projection order by order in the perturbation parameter allows us to find many linear constraints.

    Let us start with the expectation value of a charged\footnote{When we mention that the operators are charged or neutral, we refer to their R-charges, in this case $J_1$. The distinction is important for our procedure: the expectation values of charged operators lead to linear constraints, while neutral ones lead to quadratic constraints.} dimension-2 chiral primary. At zero coupling, one finds:
    \begin{equation}
        \begin{aligned}
            (N^2 - &1)^{-1}\frac{\expval{\Tr \bar X^2}{\Psi}}{\braket{\Psi}} = \\
             &-\frac{1}{2}\beta_1 + c_1 \beta_1^2 + \frac{\beta_1^3}{4 N} \qty[N(N^2 - 1) - 8 c_2 N(N^2 + 1) - 8 c_3 (2N^2 - 3) + 4 c_4 N] + \mathcal{O}(\beta_1^4)\,.
        \end{aligned}
    \end{equation}
    Notice that the parameter $c_4$ appears at the same order as $c_2$ and $c_3$ in this expression, hence the need to include it in the ansatz. We must equate this expression with the supergravity expectation value computed in \eqref{eq:vevs}:
    \begin{equation}
        (N^2 - 1)^{-1}\expval{\Tr \bar X^2}_{\text{gravity}} = -\frac{1}{2}\beta_1 - \frac{1}{3} \beta_1^3 + \mathcal{O}(\beta_1^5)
    \end{equation}
    From this we obtain $c_1 = 0$, as well as an expression for $c_4$ in terms of $c_2$ and $c_3$. A further constraint may be obtained from the expectation value of dimension-4 single particle operators, which have been argued to vanish in supergravity.
    \begin{equation}
            \frac{\expval{O_{4(4)}}{\Psi}}{\braket{\Psi}} = 
             \beta_1^2 c_3^* \frac{N^6 - 14 N^4 + 49 N^2 - 36}{N^2 + 1} + \mathcal{O}(\beta_1^3)\,,
             \label{eq:dim4matching}
    \end{equation}
    from which we conclude $c_3 = 0$. Finally we can determine $c_2$ from the expectation value of the R-charge. The free-theory expression is somewhat lengthy, depending on all the constants $c_j$, but after applying the previous relations and simplifying, we find
        \begin{equation}
            (N^2 - 1)^{-1} \frac{\expval{J_1}{\Psi}}{\braket{\Psi}} = 
             \beta_1^2 + \beta_1^4 \qty[\frac{1}{3} + \abs{8 c_ 2 - 1}^2 \frac{N^4-1}{2(N^2 - 1)}] + \mathcal{O}(\beta_1^5)\,,
    \end{equation}
    and the matching with supergravity leads to $c_2 = 1/8$. Summarizing the results, the half-BPS state is then found to be, at quadratic order:
    \begin{equation}
        \ket{\Psi} \sim 1 - \frac{1}{2} \beta_1 \Tr X^2 + \frac{1}{8} \beta_1^2 (\Tr X^2)^2 + \dots
    \end{equation}
    and one can check that all remaining equations in \eqref{eq:match_vevs} are satisfied.

    This construction and the results follow closely from \cite{Giusto:2024trt}, albeit with a different expansion parameter. Before moving on to the quarter-BPS state, we conclude this part by noting that, in the half-BPS case, it is in fact possible to construct the heavy operator for a finite value of the deformation parameter $\beta_1$ \cite{Aprile:2024lwy}. Indeed, the authors argued from the symmetries of the consistent truncation that the state must be built out of multi-traces of the form $(\Tr X^2)^k, k\geq 0$.  From this observation, one can build a simpler ansatz and then solve the simple recurrence relations that arise from the matching conditions. After resummation, one finds:
    \begin{equation}
        \ket{\Psi} \sim \exp(-\frac{1}{2}\tanh \beta_1 \Tr X^2) \,.
    \end{equation}
    
    \subsection{Identification of the quarter-BPS state}
    \label{sub:cftQuarter}

    The general procedure for identifying the quarter-BPS state is identical, but requires a more thoughtful ansatz. Naively, one could decide to include all multi-trace operators involving two scalar fields, $X$ and $Y$. However, the amount of parameters in such an ansatz is too large to be fixed by the holographic constraints. Fortunately the state is invariant under a certain number of symmetries that one can use to reduce this number.

    Indeed, note that the bulk geometry is invariant under the action of some $\mathbb{Z}_2$ subgroups of $SO(6)$, which translate to symmetries of the full quarter-BPS state:
    \begin{enumerate}
        \item 
        The geometry is invariant under reversals of the Cartesian coordinates of the 5-sphere, so the state must be invariant under
        \begin{equation}
            X \to -X \qand Y \to -Y \, \qq{(independently).}
        \end{equation}
        \item 
        Swapping the amplitudes of the two perturbations and compensating with a rotation of the 5-sphere also leaves the geometry invariant, so we find another symmetry of the CFT state:
        \begin{equation}
            (X \leftrightarrow Y \,,\ \beta_1 \leftrightarrow \beta_2) \,.
            \label{eq:sym-swap}
        \end{equation}
        \item 
        The last $\mathbb{Z}_2$ symmetry is more subtle. Consider a simultaneous $\pi/2$ rotation in each of the two planes of the 5-sphere carrying the excitations. Applying this transformation to the matrix of scalars $T_{ij}$ transforms the fields $\lambda_1 \to - \lambda_1$ and $\lambda_2 \to -\lambda_2$. This can be compensated by negating the amplitudes of the perturbation, since only odd powers of the amplitudes appear in the perturbative expansion. This is however not the only effect of the rotation, it also reverses the gauge fields. Because they only have a leg in the time direction, this can be compensated by a time reversal. On the CFT state, this corresponds to a conjugation operation. All in all, the state must be invariant under the following operation:
        \begin{equation}
            (X \to i \bar X \,,\ Y \to i \bar Y \,,\ \beta_1 \to -\beta_1 \,,\ \beta_2 \to -\beta_2 \,,\ \text{conjugation}) \,.
            \label{eq:sym-time-reversal}
        \end{equation}
    \end{enumerate}

    The multi-trace operators of dimension 2 and 4 that are compatible with these symmetries, and thus allowed in the ansatz at quadratic order, are:
    \begin{equation}
        \begin{gathered}
            \Tr X^2\,,\ \Tr Y^2\,,\ \Tr X^4\,,\ \Tr Y^4\,,\ \Tr X^2Y^2\,,\ \Tr \,(XY)^2\,,\\
            (\Tr X^2)^2\,,\ (\Tr Y^2)^2\,,\ (\Tr X^2)(\Tr Y^2)\,,\ (\Tr XY)^2 \,.
        \end{gathered} 
        \label{eq:multi-traces-no-susy}
    \end{equation}
    Not all these operators are annihilated by the same supercharges as the full CFT state. We can further reduce this list by using the explicit spectrum of quarter-BPS operators, discussed in section~\ref{sec:sym}. At dimension 4, there is a unique quarter-BPS operator respecting the symmetries, \eqref{eq:pure_quarter_state}, that we rewrite here for convenience:
    \begin{equation}
        \Op_{1/4} = \Tr X^2 \Tr Y^2 - (\Tr XY)^2 - \frac2N \qty(\Tr X^2Y^2 - \Tr \,(XY)^2) \,.
    \end{equation}
    There is also a single half-BPS operator that includes both letters $X$ and $Y$, it is the R-symmetry descendent of $(\Tr X^2)^2$:
    \begin{equation}
        O_{1/2} = \Tr X^2 \Tr Y^2 + 2 (\Tr XY)^2 \,.
    \end{equation}

    From this discussion we can now construct the following ansatz:
    \begin{equation}
        \begin{aligned}
        \ket{\Psi} ~\sim~ 1 & - \frac{1}{2} (\beta_1 \Tr X^2 + \beta_2 \Tr Y^2) \\ 
        & + \gamma_1 \Tr X^2 + \hat\gamma_1 \Tr Y^2 + \gamma_2 \Tr X^4 + \hat\gamma_2 \Tr Y^4 + \gamma_3 (\Tr X^2)^2  + \hat\gamma_3 (\Tr Y^2)^2 \\
        & + \gamma_4 O_{1/2} + \gamma_5 O_{1/4} + \dots\,.
        \end{aligned}
        \label{eq:ansatz_quarter}
    \end{equation}
    The coefficients $\gamma_i, \hat\gamma_i$ have an expansion in powers of the amplitudes of the perturbations $\beta_1, \beta_2$, starting at quadratic order:\footnote{For the same technical reasons as in section~\ref{sub:review_half_state}, we need to include the cubic terms in $\gamma_1$ and $\gamma_2$.}
    \begin{equation}
        \gamma_i = \gamma_i^{(2, 0)} \beta_1^2 + \gamma_i^{(1, 1)} \beta_1 \beta_2 + \gamma_i^{(0, 2)} \beta_2^2 + \mathcal{O}(\beta_1, \beta_2)^3 \,.
    \end{equation}
    From the symmetry \eqref{eq:sym-swap}, we can express the hatted coefficients $\hat\gamma_i$ in terms of the unhatted ones:
    \begin{equation}
        \hat\gamma_i = \gamma_i^{(0, 2)} \beta_1^2 + \gamma_i^{(1, 1)} \beta_1 \beta_2 + \gamma_i^{(2, 0)} \beta_2^2 + \mathcal{O}(\beta_1, \beta_2)^3 \,,
    \end{equation}
    and furthermore $\gamma_{4,5}^{(2, 0)} = \gamma_{4,5}^{(0, 2)}$. The symmetry \eqref{eq:sym-time-reversal}, in turn, implies that at quadratic order $\gamma_1^{(p,q)}$ are purely imaginary, while all the other coefficients are real. Remembering also that the limit $\beta_2 = 0$ must yield the half-BPS state of the previous section, we can already fix all $\gamma_i^{(2,0)}$. There remains seven unfixed coefficients\footnote{As in the half-BPS case, the cubic terms are necessary because they appear at the same order as some of the quadratic ones.}, $\gamma_{1 \leq i \leq 5}^{(1,1)}$, $\gamma_1^{(2,1)}$, and $\gamma_1^{(1,2)}$, all of them real.
    
    As we did for the half-BPS state, we will determine the coefficients in the ansatz through a matching with the holographic expectation values computed in the bulk \eqref{eq:vevs}. One has however to be careful: it is not true in general that three-point correlators involving two quarter-BPS and one half-BPS operators are protected. In fact, we find in section \ref{sub:non-protected-corr} an example where the supergravity result and the free theory result diverge. In the special case of dimension-2 chiral primaries, however, it has been established in \cite{DHoker:2001jzy} (see also \cite{Bissi:2021hjk}), that all three-point functions $\expval{O_{pq} O_{rs} O_2}$ are indeed protected (where $p,q$ and $r,s$ denote the representations of the quarter-BPS operators). This property is enough to ensure that the following matching holds at quadratic order:
    \begin{equation}
        \frac{\expval{\Op_{2}}{\Psi}}{\braket{\Psi}} \ \hat=\ \expval{\Op_{2}}_{\text{sugra}} \,,\quad
        \frac{\expval{J_{1,2}}{\Psi}}{\braket{\Psi}} \ \hat=\ \expval{J_{1,2}}_{\text{sugra}}% \,,\quad \frac{\expval{J_2}{\Psi}}{\braket{\Psi}}  = \expval{J_2}_{\text{gravity}}\,,
        \label{eq:quarter_matching}
    \end{equation}
    where $O_2$ is a dimension-2 chiral primary.

    When $O_2 = \Tr \bar X^2$, we compute
    \begin{equation}
        \begin{aligned}
            (N^2 - 1)^{-1}&\frac{\expval{\Tr \bar X^2}{\Psi}}{\braket{\Psi}} =
             -\frac{1}{2}\beta_1 - \gamma_1^{(1,1)}\beta_1\beta_2 + \frac{1}{3}\beta_1^3 \\
             &  \quad + \qty(-\gamma_1^{(2,1)} + 2(N^2 + 1) \gamma_3^{(1,1)} + \frac{4N^2 - 6}{N} \gamma_2^{(1,1)})\beta_1^2\beta_2 \\
             &  \quad + \qty(- \frac{N^2- 1}{4} -\gamma_1^{(1,2)} + (N^2 + 1) \gamma_4^{(1,1)} + (N^2 - 4) \gamma_5^{(1,1)})\beta_1\beta_2^2 \\
             &  \quad + \mathcal{O}(\beta_1,\beta_2)^4\,.
        \end{aligned}
    \end{equation}
    from which equating with \eqref{eq:vevs} leads to $\gamma_1^{(1,1)} = 0$ as well as expressions for $\gamma_1^{(1,2)}$ and $\gamma_1^{(2,1)}$.
    
    We then turn to the matching of the R-charges. Remarkably, we find that the resulting equation is a positive definite quadratic form, that can be written as a sum of squares with positive coefficients (when $N$ is sufficiently large):
    \begin{equation}
        \begin{aligned}
        0 =& \frac{4(N^4 - 6N^2 + 18)}{N^2}\abs{\gamma_2^{(1,1)} + \frac{2N(2N^2 - 3)}{N^4 - 6N^2 + 18} \gamma_3^{(1,1)}}^2 + \frac{8N^2(N^2 - 4)(N^2 - 9)}{N^4 - 6N^2 + 18} \abs{\gamma_3^{(1,1)}}^2 \\
        &+ 6 (N^2+1) \abs{\gamma_4^{(1,1)} - \frac{1}{12}}^2 + 3 (N^2 - 4)\abs{\gamma_5^{(1,1)} - \frac{1}{6}}^2 \,.
        \end{aligned}
    \end{equation} 
    The solution to this equation is unique, and fixes all remaining coefficients: $\gamma_2^{(1,1)} = \gamma_3^{(1,1)} = 0$, $\gamma_4^{(1,1)} = 1/12$, and $\gamma_5^{(1,1)} = 1/6$.
    The quarter-BPS state dual to the bulk geometry is threfore found to be:
    \begin{equation}
		\begin{aligned}
			\ket{\psi} \ &\sim\ 1 - \frac12 \qty(\beta_1 \Tr X^2 + \beta_2 \Tr Y^2) + \frac1{8} \qty(\beta_1^2 (\Tr X^2)^2 + \beta_2^2 (\Tr Y^2)^2) \\
            &\qquad + \frac{1}{12} \beta_1 \beta_2 (O_{1/2} + 2 O_{1/4})  + \bigO\qty(\beta_1,\beta_2)^3 \,.
		\end{aligned}
		\label{eq:heavy_state_q}
	\end{equation}
    This proposed solution is also compatible with the expectation values of all other dimension-2 operators computed in supergravity.

    In the next section we will see a consequence of this identification on the expectation values of higher-dimensional operators. Before that, we wish to make some comments on the analysis performed here. First, we highlight that the ansatz \eqref{eq:ansatz_quarter} has been constructed with the knowledge of the precise from of the quarter-BPS operator \eqref{eq:pure_quarter_state}, and this ingredient is in fact essential to the computation. Without it, one would build a more general ansatz, including a term of the form
    \begin{equation}
        \gamma_6 (\Tr XXYY - \Tr XYXY) \,.
    \end{equation}
    However, the quadratic constraint derived from the R-charge matching is no longer positive definite in that case, and so the solution would not be unique.

    Note also that we have used the identification of the half-BPS state to fix some coefficients of the ansatz. Again, without this knowledge, it would not have been possible to completely fix the state. The reason is that the identification of the half-BPS state relies on the matching of the expectation values of dimension-4 chiral primaries \eqref{eq:dim4matching} ; these are not necessarily protected in the quarter-BPS case.

    \subsection{A non-protected quarter-BPS correlator}
    \label{sub:non-protected-corr}

    In this section we investigate three-point functions of the type:
    \begin{equation}
        \expval{ \bar O_{1/4} O_4 O_{1/4} }
    \end{equation}
    where $O_{1/4}$ is defined in \eqref{eq:pure_quarter_state} and $O_4$ is a half-BPS operator of dimension 4. These correlators have been shown to receive no corrections at one loop \cite{DHoker:2001jzy}, and are expected to be protected at all orders. Thanks to the identification to quadratic order of the quarter-BPS heavy state \eqref{eq:heavy_state_q}, we can compute these correlators in the supergravity regime. We will find that one such correlator differs from its free-theory value, and is therefore not protected.

    In section~\ref{sub:holography}, we argued that the expectation values of single-particle operators of dimension greater than 2 vanish in the quarter-BPS background. Indeed, the geometries belong to a consistent truncation that projects out the scalar fields not transforming in the $[0, 2, 0]$ representation. The asymptotics of these fields are directly mapped through the holographic dictionary \cite{Turton:2025svk}, to the expectation values of single-particle operators. In particular, we find
    \begin{equation}
        \expval{O_{4,I}}_{\text{sugra}} = 0 \,,
    \end{equation}
    where $I$ labels the single-particle operators of dimension 4.

    Consider as a first example the single-particle operator of highest-weight,
    \begin{equation}
        O_{4(4)} = \Tr X^4 - \frac{2 N^2-3}{N (N^2+1)} (\Tr X^2)^2 \,.
    \end{equation}
    We can compute the expectation value in the free theory using Wick contractions and the form of the state \eqref{eq:heavy_state_q}, finding
    \begin{equation}
        \expval{O_{4(4)}}{\Psi}_{\text{free}} = 0 \,.
    \end{equation}
    This matches with the supergravity result. This is no surprise: after expanding the heavy state, most of the three-point functions entering the computation have been shown to be protected. The only ones for which no such proof exists involve two quarter-BPS operators and a half-BPS operator; in the present case that would be $\expval{\bar O_{1/4} O_{4(4)} O_{1/4}}$, but this correlator trivially vanishes due to R-charge conservation.

    The same reasoning applies for the expectation values of all dimension-4 single-particle primaries charged under R-symmetry, and indeed, we have checked that the expectation values of these operators in supergravity match with the free-theory values.

    We therefore turn to the non-trivial cases, in which the single-particle operator is neutral. These operators can be constructed by acting with the $SO(6)$ lowering operators on the highest-weight state, see \cite{Turton:2025svk} for a review of this procedure. In all cases but one, we again find agreement between supergravity and free-theory results. However, there is a discrepancy for one operator. Defining the following lowering operators
    \begin{equation}
        E_1 = \bar X \pdv{\bar Z} - Z \pdv{X} \qand E_2 = \bar X \pdv{Z} - \bar Z \pdv{X} \,,
    \end{equation}
    the operator in question is
    \begin{equation}
        O_{4(0)} =~ E_1^2 E_2^2 O_{4(4)}
        \label{eq:O40}
    \end{equation}
    and its expression in terms of the fundamental fields can be found in Appendix~\ref{app:spOp4}.
    Note that this operator is not normalized, as it is not necessary for our purposes.
    Computing its expectation value in the free theory, we find:
    \begin{equation}
        \expval{O_{4(0)}}{\psi}_{\text{free}} = \frac{16 (N^2 - 1)(N^2 - 4)(N^2 - 9)}{9N(N^2 + 1)} \beta_1^2 \beta_2^2 + \mathcal{O}(\beta_1, \beta_2)^5 \,.
        \label{eq:discrepancy_ev}
    \end{equation}
    This differs from the supergravity result, which vanishes. Note that although we do not have access to the cubic (or higher-order) terms of the state $\ket{\psi}$, these terms would only contribute to \eqref{eq:discrepancy_ev} starting at order 5. Indeed, the contributions of order 4 vanish either by R-charge conservation, or because single-particle operators are orthogonal to double-traces.
    After stripping away the contributions that are known to be protected, we find the following discrepancy
    \begin{align}
        \expval{\bar O_{1/4} O_{4(0)} O_{1/4}}_{\text{sugra}} &= 0 \,,\\
        \expval{\bar O_{1/4} O_{4(0)} O_{1/4}}_{\text{free}} \ &= \frac{64 (N^2 - 1)(N^2 - 4)(N^2 - 9)}{N(N^2 + 1)} \,,
    \end{align}
    thereby showing that despite receiving no corrections at one-loop, this correlator is in fact not protected.

    \section{Numerical results}
    \label{sec:numerics}
    
    As already stated, the supergravity results in this work have also been checked by solving the full supergravity equations from Sec.~\ref{sub:BPSeqs} numerically, and this section provides the details.
    
    While we can solve a set of decoupled, second-order differential equations for $\sigma_{\pm}$, \eqref{eq:decoupled_bps}, in practice it is much easier, and faster in terms of convergence, to take advantage of the perturbation theory and apply a relaxation method to the first order BPS equations: \eqref{eq:mu_BPS}, \eqref{eq:omega0_bps}, \eqref{eq:phi_BPS}, \eqref{eq:lambda_bps} and \eqref{eq:integral_BPS}. In the actual code we replace the last equation, \eqref{eq:integral_BPS}, with a differential one (from which the integral of motion can be derived):
    \begin{equation}
    	\xi\,\partial_\xi\,\log\,G=-2\,(G-1)(1-\phi_1\,\cosh\,\lambda_1-\phi_2\,\cosh\,\lambda_2).
    \end{equation}
    The algebraic equations can, of course, be used to reduce the number of unknowns we need to integrate and we will use them to eliminate the gauge fields, $\phi_{1,2}$ and the scalar $\mu_1$. Hence, we solve numerically for $\{\mu_2,\,\lambda_1,\,\lambda_2,\,\omega_0,\,\omega_1\}$ as functions of the radial variable $\xi\in[0,1]$. The numerical integration is carried out by using spectral collocation methods on a Chebyshev-Gauss-Lobatto grid and implementing a standard Newton-Raphson iteration (see \cite{Dias:2015nua} for a review of such methods applied to gravity). We need to impose appropriate boundary conditions to ensure a well-posed problem.
    
    At the origin, $\xi=0$, regularity enforces Neumann boundary conditions on almost all fields we are interested in, except for $\lambda_{1,2}$:
    \begin{equation}
    	\left.\partial_\xi\,X(\xi)\right|_{\xi=0}=0,\quad{\rm for}\,X\in\{\mu_2,\,\omega_0,\,\omega_1\}.
    \end{equation}
    Our solutions depend on 2 parameters -- the amplitudes of $\lambda_{1,2}$, and following the perturbation theory normalization, \eqref{eq:lambda12Norm}, we will fix them at the origin via:
    \begin{equation}
    	\lambda_{1,2}(\xi=0)=2\,\beta_{1,2}.
    \end{equation}
    The spacetime is asymptotically locally AdS and at the AdS boundary, $\xi=1$, we demand that it approaches the global AdS$_5$ vacuum, which translates in the following conditions for the functions we will solve for:
    \begin{equation}
    	\mu_2(\xi=1)=\lambda_{1,2}(\xi=1)=0,\quad\omega_0(\xi=1)=\omega_1(\xi=1)=1.
    \end{equation}
    
  	The space of geometries is explored by using the perturbative solution as a seed to the Newton-Raphson method. We start from $\beta_{1,2}$ very small ($\sim0.1$) and gradually increase their values, using previous solutions as new seeds. The case of $\beta_1=\beta_2$ is slightly simpler with faster convergence for values of $\beta_1=\beta_2>1$. All the solutions presented in this work have been checked to satisfy the equations of motion to $10^{-50}$ when varying the grid size.
  	
\begin{figure*}[ht]
  		\centering
  		\subfigure{\includegraphics[width=.98\textwidth]{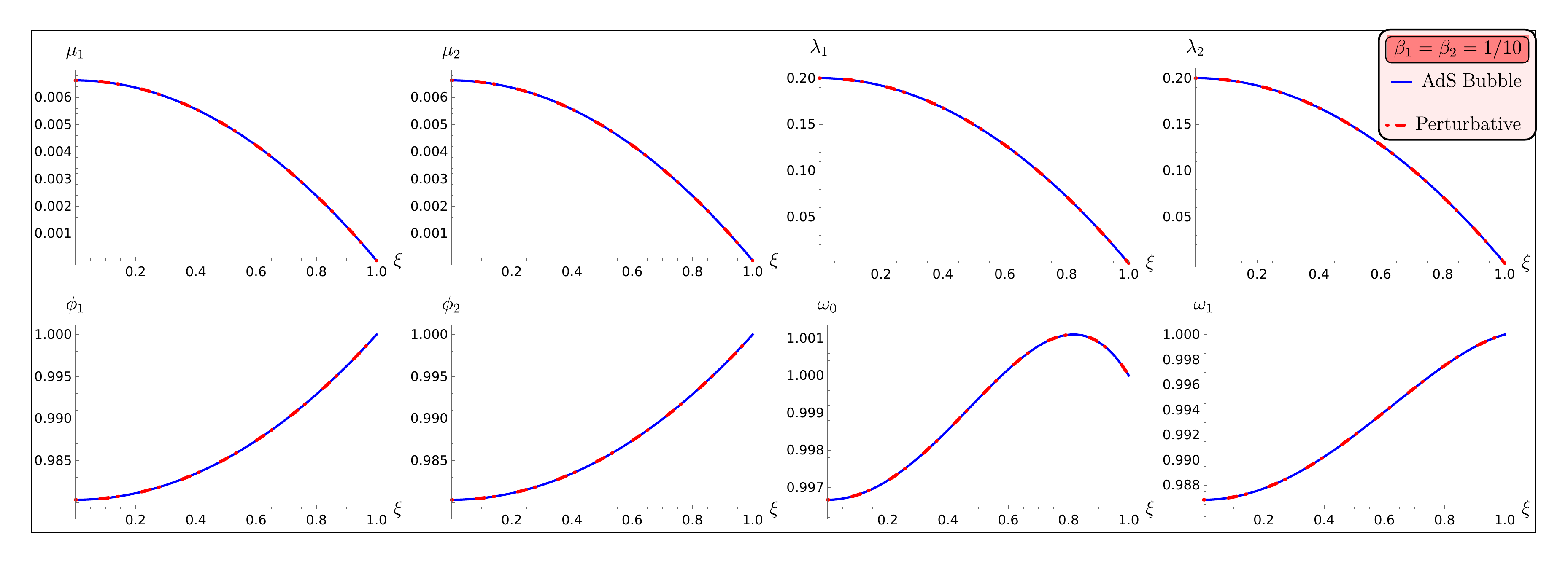}
  		}
  		\caption{Comparison between the full numerical solutions (solid blue lines) and the perturbative results at order 12 (dash-dotted red lines) for $\beta_1=\beta_2=1/10$. With these small values, the perturbation theory is a very good approximation of the real solutions.}
  		\label{fig:SolsComp1}
  	\end{figure*}

\begin{figure*}[ht]
  		\centering
  		\subfigure{\includegraphics[width=.98\textwidth]{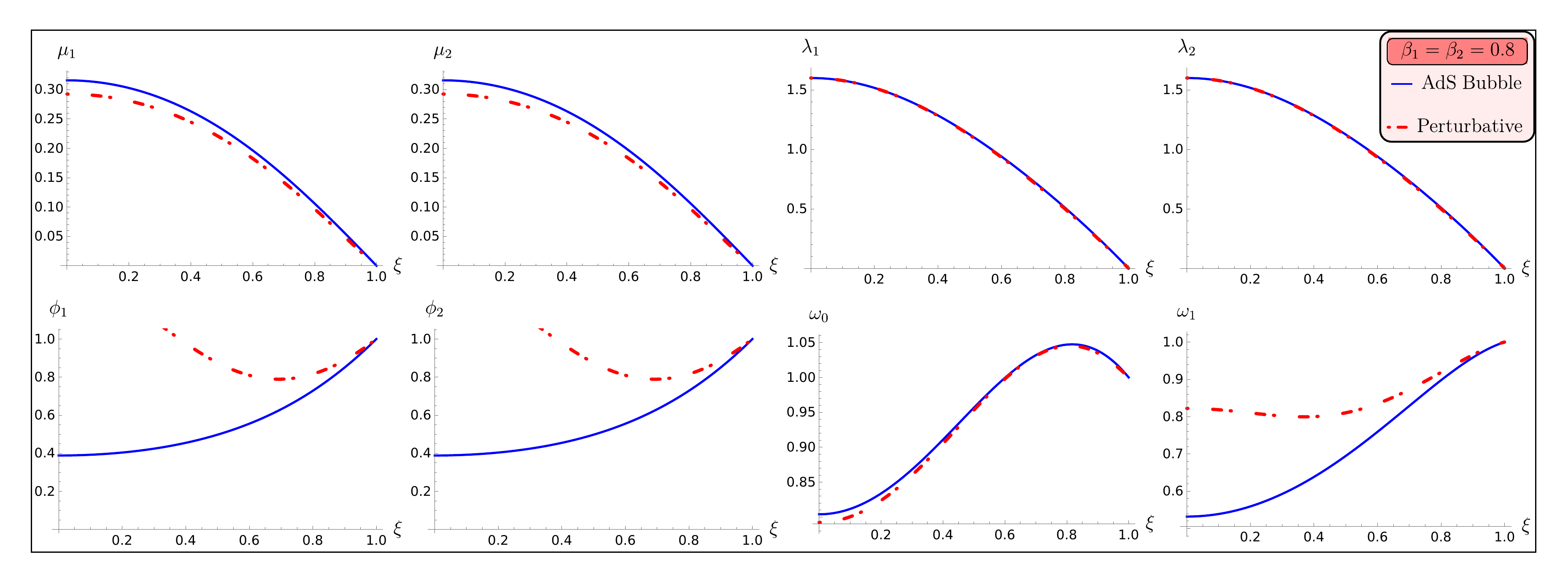}
  		}
  		\caption{Comparison between the full numerical solutions (solid blue lines) and the perturbative results at order 12 (dash-dotted red lines) for $\beta_1=\beta_2=8/10$. At these values of the deformation parameters, the perturbation theory starts to deviate significantly from the numerical solutions.}
  		\label{fig:SolsComp2}
  	\end{figure*}

\begin{figure*}[ht]
  		\centering
  		\subfigure{\includegraphics[width=.9\textwidth]{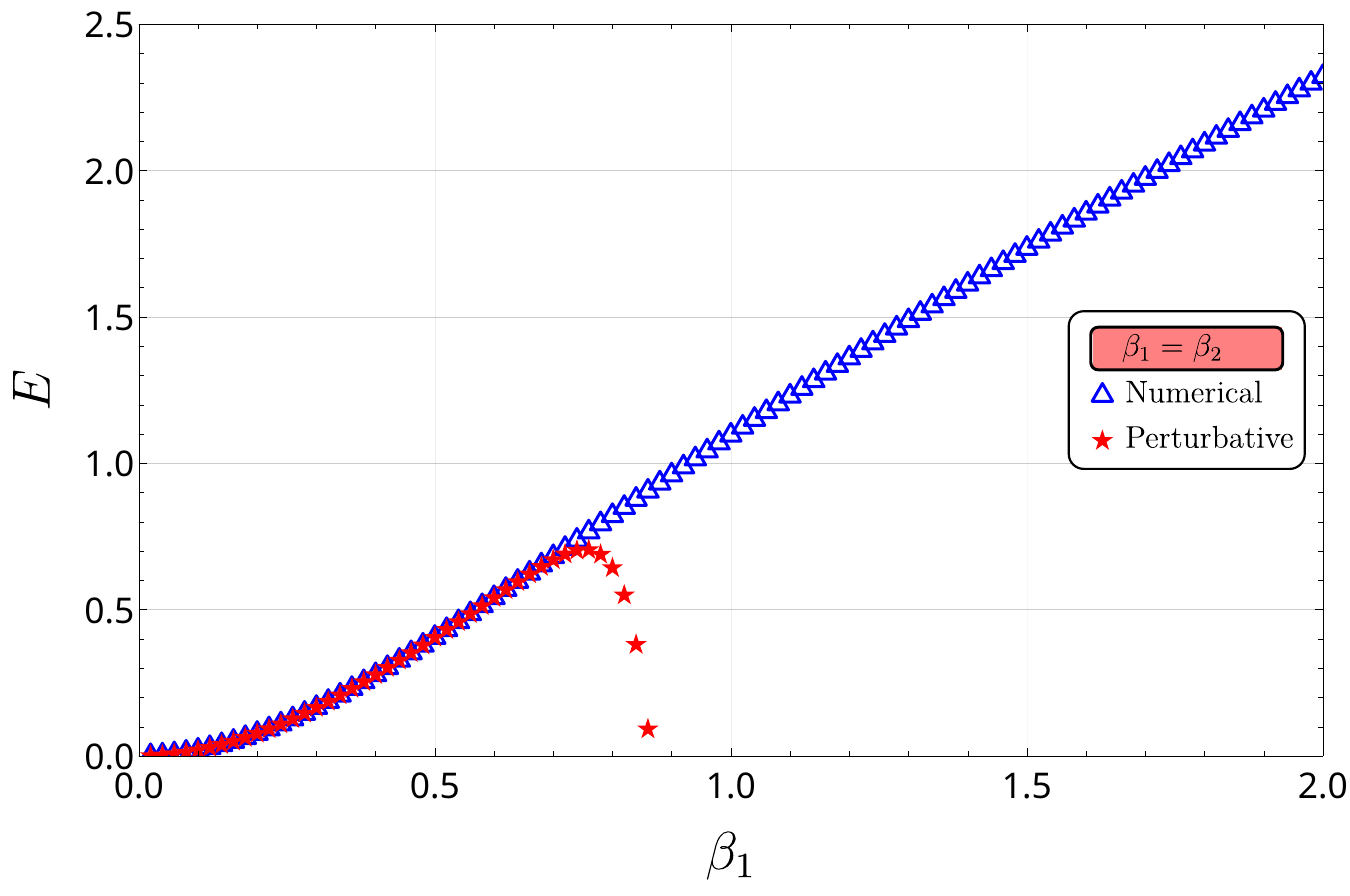}
  		}
  		\caption{Energy, \eqref{eq:energy}, of the solution, for equal $\beta_1=\beta_2$ with the full numerical result (blue hollow triangles) in comparison to the perturbative results at 12th order (red filled stars).}
  		\label{fig:EnEqualBeta2}
  	\end{figure*}

\begin{figure*}[ht]
  		\centering
  		\subfigure{\includegraphics[width=.98\textwidth]{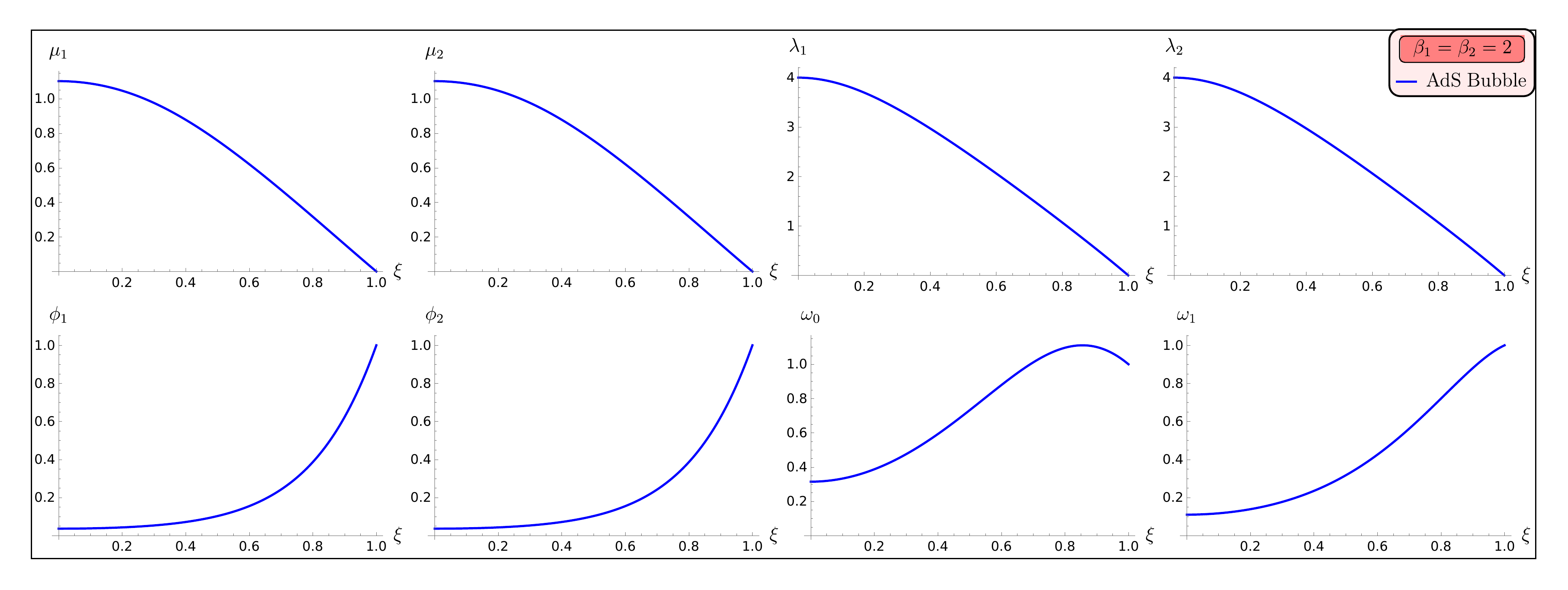}
  		}
  		\caption{The full numerical solutions for $\beta_1=\beta_2=2$. The perturbative results are not plotted, as they diverge too much from the numerical solutions for these large deformation parameters.}
  		\label{fig:SolsComp3}
  	\end{figure*}

  	For small values of $\beta_{1,2}$ we find that the perturbative solution (dash-dotted red lines on all plots) agrees spectacularly well with the numerical results (blue solid lines), as seen in Fig.~\eqref{fig:SolsComp1} for $\beta_1=\beta_2=1/10$. The perturbative results start diverging around $\beta_1=\beta_2=0.8$, shown in Fig.~\eqref{fig:SolsComp2}, and for $\beta_1=\beta_2\sim1$ they no longer provide a good approximation, even as a seed solution for the numerical integration. This is better illustrated on a plot of the energy of the solution, \eqref{eq:energy}, computed using the holographic stress-energy tensor\footnote{Remember that we have set the AdS radius to 1, hence all quantities are measured in units of $R_{\rm AdS}$.}, as a function of $\beta_1$, Fig.~\eqref{fig:EnEqualBeta2}. Fig.~\eqref{fig:SolsComp3} presents the case of $\beta_1=\beta_2=2$ where the perturbation theory has been fully omitted because it only matches with the numerics at the boundaries, due to the imposition of boundary conditions. We have gone all the way to $\beta_1=\beta_2=5$ in the case of equal betas, but there is very little qualitative difference in the results. In practice the only limitation is the amount of computer resources needed to reach even higher values.
  	
  	
\begin{figure*}[ht]
  		\centering
  		\subfigure{\includegraphics[width=.98\textwidth]{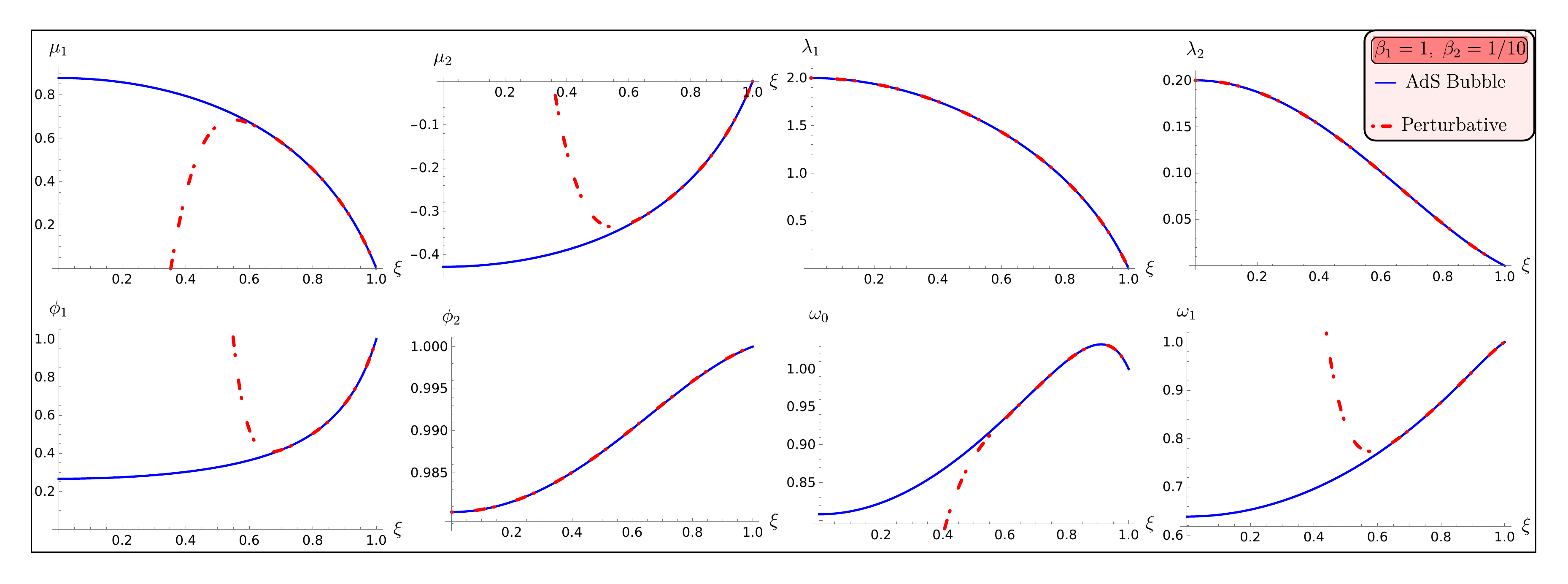}
  		}
  		\caption{Comparison between the full numerical solutions (solid blue lines) and the perturbative results (dash-dotted red lines) at order 12 for $\beta_1=1$ and $\beta_2=1/10$.}
  		\label{fig:SolsComp4}
  	\end{figure*}

  	We also explored the part of the phase space where the deformation parameters are not equal. In that case, one loses the symmetry between the two sectors -- that is, $\mu_1$, $\lambda_1$ and $\phi_1$ differ from $\mu_2$, $\lambda_2$ and $\phi_2$. Nevertheless, exchanging the values of $\beta_{1,2}$ is equivalent to $\{\mu_1,\,\lambda_1\,,\phi_1\}\leftrightarrow\{\mu_2,\,\lambda_2\,,\phi_2\}$, allowing us to cut in half the parameter space under consideration. We provide an illustrative example, where the perturbative results are still useful, for $\beta_1=1$ and $\beta_2=1/10$ in Fig.~\eqref{fig:SolsComp4}. The larger the difference between $\beta_1$ and $\beta_2$, the larger the gradients encountered in the numerics become, requiring more precision and, thus, a denser grid. We have verified our results within the whole space $\beta_{1,2}\in[1/10,\,2]$.
  	
  	Noticeable, especially from Fig.~\eqref{fig:SolsComp4}, is that the perturbation theory is a better approximation near the asymptotic boundary of the spacetime. This can be explained with the weaker effects of backreaction far from the AdS bubble centered at the origin, combined with the fact that the geometry approaches vacuum AdS$_5$ asymptotically, which is also the background around which we expand perturbatively in Sec.~\eqref{sub:perturbations}. 
  	
\begin{figure*}[ht]
  		\centering
  		\subfigure{\includegraphics[width=0.45\textwidth]{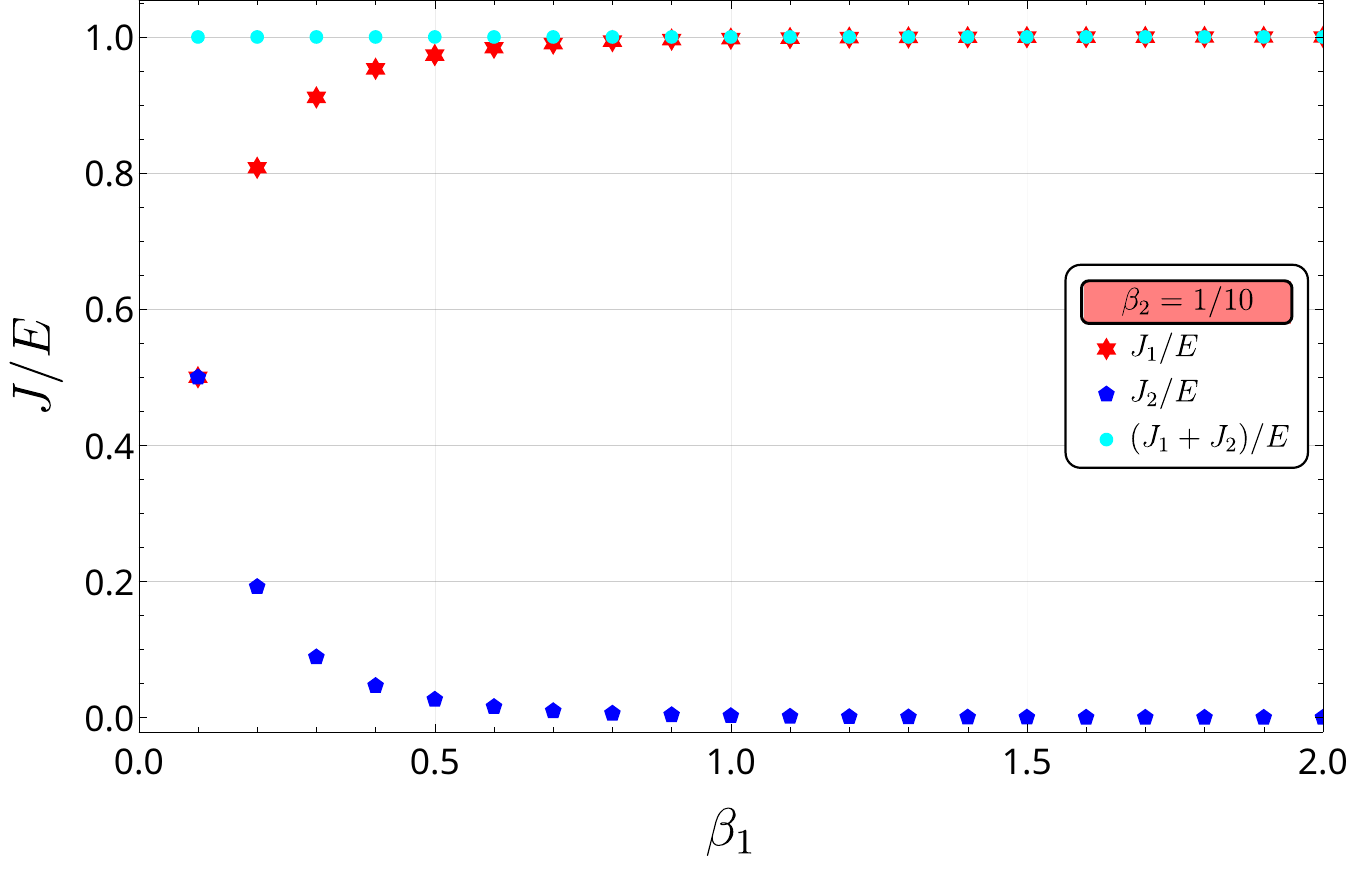}
  		}
  		\subfigure{\includegraphics[width=0.45\textwidth]{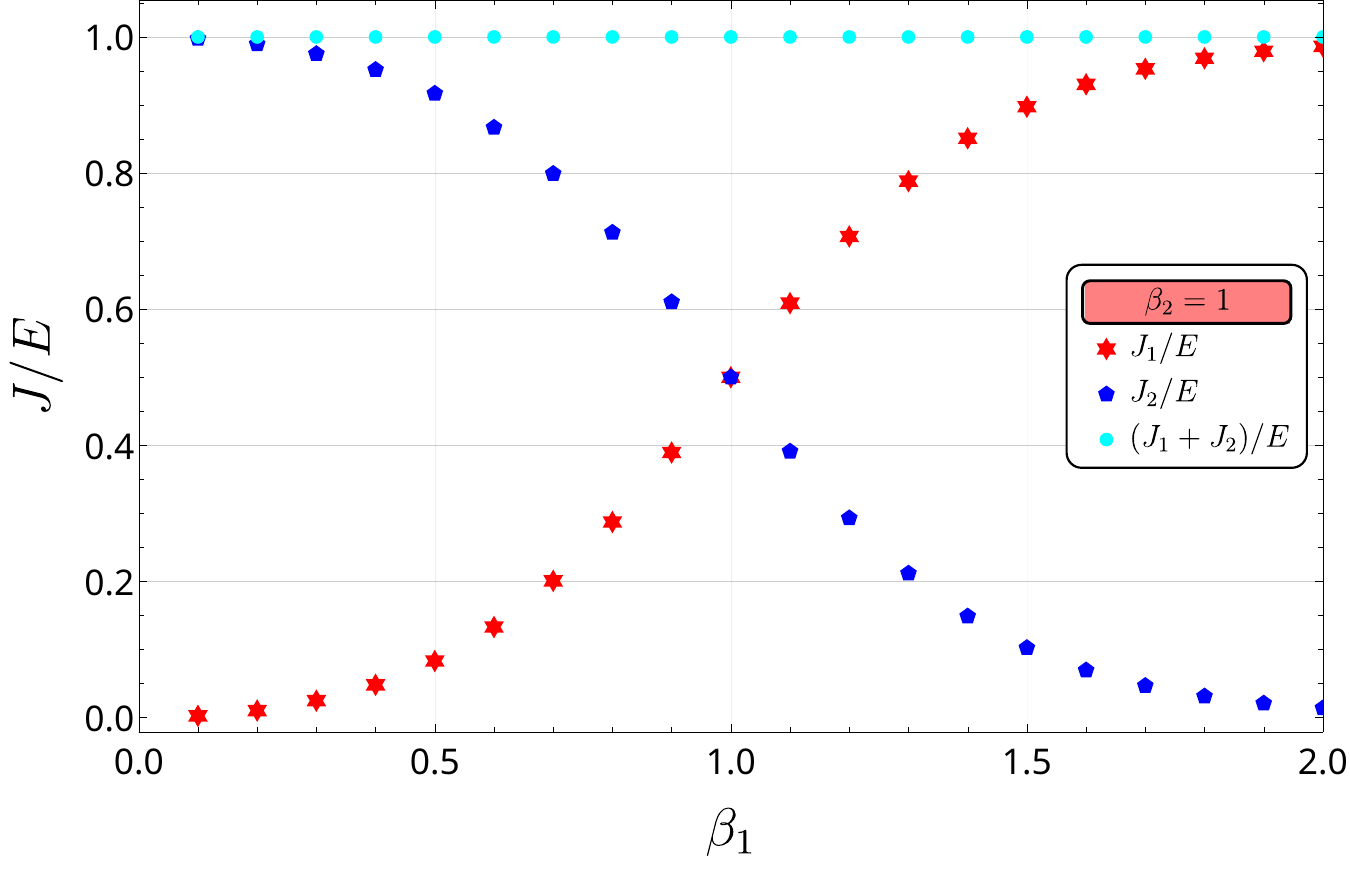}
  		}
  		\caption{Ratio of the holographic R-charges $J_{1,2}$, \eqref{eq:charges}, to the energy of the geometry, \eqref{eq:energy}, individually and as a sum, as a function of $\beta_1$ for fixed $\beta_2$. All our geometries verify the BPS condition $E=J_1+J_2$, as confirmed by the green circles and extending way beyond the range of validity of the perturbation theory.}
  		\label{fig:JEComp}
  	\end{figure*}

  	Finally, as a consistency check of our numerical scheme we compute the R-charges and the energy of the geometries using the holographic formulas \eqref{eq:charges}, \eqref{eq:energy}. These solutions are supersymmetric, and we expect them to verify the BPS condition $E=J_1+J_2$, and this is indeed what we find, as demonstrated in Fig.~\eqref{fig:JEComp}. We also present the ratios $J_{1,2}/E$. Unsurprisingly they are correlated with the relative sizes of $\beta_{1,2}$ and, when $\beta_1=\beta_2$, the two R-charges are equal $J_1 = J_2$.
  	
    \section{Discussion}
    \label{sec:conclusion}

    In this work, we have presented a concrete realization of the AdS$_5$/CFT$_4$ dictionary for quarter-BPS heavy states. Our analysis focused on a specific class of quarter-BPS geometries, known as \ads{} bubbles, that fit within a five-dimensional truncation of Type IIB supergravity. With a suitable ansatz, we were able to decouple the BPS equations governing these solutions, allowing us to better study the asymptotic behavior of the solutions. We examined these geometries both perturbatively, in an expansion around the \adsf{} vacuum, and numerically, solving the full non-linear equations.

    We then performed a holographic analysis of the bubbling solutions. From an asymptotic expansion of the supergravity fields, we computed the expectation values of all light operators of dimension-2, as well as the R-charges, perturbatively in terms of the deformation parameters of the solutions, $\beta_1$ and $\beta_2$.
    We constructed an ansatz for the dual CFT state to quadratic order in the perturbation parameter, and employed a non-renormalization theorem to match the three-point functions computed in the free theory with the holographic results. This allowed us to determine the coefficients of the ansatz, and finally express the CFT state in terms of the lightest quarter-BPS operator \eqref{eq:pure_quarter_state}. 
    
    Subsequently, we used this identification to compute quarter-BPS correlators at strong coupling. These correlators were shown to not receive corrections at one loop, and were expected to be protected at all orders. We found however a counter-example: one instance where the supergravity result differs from the free-theory result, therefore showing that this correlator is not protected.

    This work represents opens avenues for further research on quarter-BPS states and geometries. One direction currently under investigation \cite{placeholder2026} involves using these geometries as a tool to compute new classes of four-point correlators with quarter-BPS operators. Indeed, according to the holographic dictionary, Heavy-Heavy-Light-Light (HHLL) correlators can be expressed in terms of the solutions to a wave equation in the geometry dual to the heavy state. Following the approach of \cite{Aprile:2024lwy,Aprile:2025hlt} (and building on earlier work in AdS$_3$ \cite{Ceplak:2021wzz}), one can consider the formal light limit of these correlators, to extract LLLL correlators. Using the expression for the CFT state derived in this paper, it becomes possible to isolate the pure quarter-BPS contribution, and compute for the first time correlators such as $\expval{O_{1/4}O_{1/4}O_{1/2}O_{1/2}}$ and $\expval{O_{1/4}O_{1/2}O_{1/2}O_{1/2}}$, in the supergravity regime.

    A natural question for future exploration is whether one can derive an expression for the heavy state beyond the perturbative regime. The expectation values of all single-particle operators of dimension higher than 2 vanish, which places strong constraints on the form of the heavy state. Combined with the expansion of the expectation values of dimension-two chiral primaries, this data may provide sufficient information to completely fix the state. The first challenge is in determining a complete, suitable ansatz in terms of multi-traces, or multi-particles, for the CFT state. Another possibility involve starting with a coherent ansatz such as \eqref{eq:quarter_coherent}, though the three-point functions of such operators are technically challenging to compute.

    Another promising direction is to explore other geometries within the consistent truncation. By adding angular momentum to the half-BPS bubble, one can construct spinning quarter-BPS solutions. The supergravity solutions are more involved, as the fields acquire a non-trivial angular dependence. At the linear level, the CFT duals are descendants of the lightest chiral primaries. A preliminary analysis of the quantum numbers at quadratic order, based on the classification in  \cite{Dolan:2002zh}, suggests that these states may belong to semi-short representations of the superconformal group. The truncation also contains 1/8- and 1/16-BPS solutions, the latter being of great interest to the study of black hole microstates. It would be interesting to perform the similar holographic analyses for these geometries, potentially revealing new insights into the microscopic structure of black holes.

	\acknowledgments
    
    We thank Stefano Giusto and Rodolfo Russo for initial collaboration, insightful discussions and comments on a draft of this paper. We also thank F. Aprile, F. Coronado, A. Holguin, J. Vilas Boas and N. Warner for valuable exchanges.
    The work of BG is supported in part by the Simons Collaboration on Global Categorical Symmetries and also by the NSF grant PHY- 2412361. AH is supported by a grant from the Swiss National Science Foundation, as well as via the NCCR SwissMAP.

	%%%%%%%%%%%%%%%%%%%%%%%%%%%%%%%%%%%%%%%%%%%%%%%
	\appendix
	%%%%%%%%%%%%%%%%%%%%%%%%%%%%%%%%%%%%%%%%%%%%%%%
		
    \section{Spherical harmonics and chiral primaries}
    \label{app:spherical_harmonics}

    In this Appendix we present the spherical harmonics of $S^5$ and some of their properties, following the conventions of \cite{Skenderis:2007yb}. 
    
    \subsection{Spherical harmonics}

    The scalar spherical harmonics are defined by
    \begin{equation}
        \Box_{S^5} \, Y^{k,I} = \Lambda_k \, Y^{k, I} \,, \quad \Lambda_k = -k(k+4) \,.
    \end{equation}
    where $k$ is the rank of the harmonics and $I$ labels the $SO(6)$ quantum numbers.
    In terms of Cartesian coordinates of the sphere $(y_i)_{i=1\dots 6}$, $y^i y^i = 1$, they can be written as
    \begin{equation}
        Y^{k,I} = C^{I}_{i_1\dots i_k} y^{i_1} \dots y^{i_k} \,.
    \end{equation}
    where $C^{I}$ are rank-$k$ totally symmetric traceless tensors.
    There is a one-to-one map between the scalar harmonics and the single trace chiral primary operators  \eqref{eq:single_trace_chirals}, defined as
    \begin{equation}
        \mathcal{T}_{k,I} = C^I_{i_1 \dots i_k} \Tr(\Phi^{i_1} \dots \Phi^{i_k}) \,.
    \end{equation}
    The normalization of the harmonics is fixed by
    \begin{equation}
        \int_{S^5} Y^{k, I_1} \overline{Y^{k, I_2}} = \pi^3 z(k) \delta^{I_1 I_2} \,, \quad z(k) = \frac{1}{2^{k-1}(k+1)(k+2)} \,.
    \end{equation}

    Let the metric of the 5-sphere be:
    \begin{equation}
        ds_{S^5}^2 = d\theta^2 + \sin^2\theta\, d\Phi_1^2 + \cos^2 \theta \qty(d \eta^2 + \sin^2 \eta \,d\Phi_2^2 + \cos^2 \eta \, d\Phi_3^2) \,.
    \end{equation}
    We are mostly interested in scalar harmonics with $k=2$, and we present a basis for the 10 independent $SO(2)$-invariant\footnote{In the bulk, this invariance corresponds to translations of $\Phi_3$. In the CFT, it is translations of the phase of the complex scalar $Z$.} harmonics, along with their corresponding single-trace chiral primary:
    {\allowdisplaybreaks
    \begin{alignat}{2}
        Y^{2, \underline{1}} &= \frac{1}{2} \sin^2\theta \,e^{2 i \Phi_1} &\quad\leftrightarrow\quad \ \mathcal{T}_{2,\underline{1}} &=  \Tr X^2 \,,\\
        Y^{2, \underline{2}} &= \frac{1}{2} \sin^2\theta \,e^{-2 i \Phi_1} &\quad\leftrightarrow\quad\  \mathcal{T}_{2,\underline{2}} &=  \Tr \bar X^2 \,,\\
        Y^{2, \underline{3}} &= \frac{1}{4\sqrt{3}} (1- 3\cos 2\theta) &\quad\leftrightarrow\quad\  \mathcal{T}_{2,\underline{3}} &= \frac{1}{\sqrt{3}} \Tr (2 X \bar X - Y \bar Y - Z \bar Z) \,,
        \\
        Y^{2, \underline{4}} &= \frac{1}{2} \cos^2\theta \sin^2\eta \,e^{2 i \Phi_2} &\quad\leftrightarrow\quad\  \mathcal{T}_{2,\underline{4}} &=  \Tr Y^2 \,,\\
        Y^{2, \underline{5}} &= \frac{1}{2} \cos^2\theta \sin^2\eta \,e^{-2 i \Phi_2} &\quad\leftrightarrow\quad\  \mathcal{T}_{2,\underline{5}} &=  \Tr \bar Y^2 \,,\\
        Y^{2, \underline{6}} &= -\frac{1}{2} \cos^2\theta \cos 2\eta &\quad\leftrightarrow\quad\  \mathcal{T}_{2,\underline{6}} &=  \Tr (Y \bar Y - Z \bar Z) \,,
        \\
        Y^{2, \underline{7}} &= \frac{1}{2 \sqrt{2}} \sin 2\theta \sin \eta\, e^{i(\Phi_1 +\Phi_2)} &\quad\leftrightarrow\quad\  \mathcal{T}_{2,\underline{7}} &= \sqrt{2} \Tr XY \,,\\
        Y^{2, \underline{8}} &= \frac{1}{2 \sqrt{2}} \sin 2\theta \sin \eta\, e^{-i(\Phi_1 +\Phi_2)} &\quad\leftrightarrow\quad\  \mathcal{T}_{2,\underline{8}} &= \sqrt{2} \Tr \bar X \bar Y \,,
        \\
        Y^{2, \underline{9}} &= \frac{1}{2\sqrt{2}} \sin 2\theta \sin \eta\, e^{i(\Phi_1 -\Phi_2)} &\quad\leftrightarrow\quad\  \mathcal{T}_{2,\underline{9}} &=  \sqrt{2} \Tr X \bar Y \,,\\
        Y^{2, \underline{10}} &= \frac{1}{2 \sqrt{2}} \sin 2\theta \sin \eta\, e^{-i(\Phi_1 -\Phi_2)} &\quad\leftrightarrow\quad \mathcal{T}_{2,\underline{10}} &=  \sqrt{2} \Tr \bar XY \,.
    \end{alignat}
    }
    The first three harmonics form the standard basis for $SO(4)$-invariant harmonics, and we have completed it to a full basis of $SO(2)$-invariant harmonics.
    
    We also define the vector and symmetric-tensor $S^5$ harmonics as the solutions of:
    \begin{equation}
        \begin{aligned}
            \Box_{S^5} \, Y_a^{k,I_5} &= \Lambda^{(v)}_k \, Y_a^{k, I_5} \,, \quad \Lambda^{(v)}_k = -(k^2 + 4k - 1) \,,
            \\
            \Box_{S^5} \, Y_{(ab)}^{k,I_{14}} &= \Lambda^{(t)}_k \, Y_{(ab)}^{k, I_{14}} \,, \quad \Lambda^{(t)}_k = -(k^2 + 4k - 2) \,,
        \end{aligned}
    \end{equation}
    along with the orthogonality conditions
    \begin{equation}
        D^a Y_a^{k, I_5} = D^a Y_{(ab)}^{k, I_{14}} = 0 \,.
    \end{equation}
    Of interest to us are the vector harmonics of rank $k=1$, used to compute the R-charges holographically in \eqref{eq:charges}. They are given by
    \begin{equation}
        \begin{aligned}
            Y_a^{1,\underline{1}} \,dy^a &= \frac{1}{\sqrt{2}} \sin^2 \theta \,d\Phi_1 \,,\\
            Y_a^{1,\underline{2}} \,dy^a &= \frac{1}{\sqrt{2}} \cos^2 \theta \sin^2 \eta \,d\Phi_2 \,.
        \end{aligned}
    \end{equation}
    where the normalization has been fixed by
    \begin{equation}
        \int_{S^5} g^{(0)ab} Y_a^{k,I_1} \overline{Y_b^{k,I_2}} = \pi^3 z(k) \delta^{I_1 I_2}\,.
    \end{equation}

    \subsection{Some three-point correlators of chiral primaries}
    \label{app:correlators}

    The identification of the heavy state presented in this article, requires us to compute a certain number of three-point functions at zero coupling, using Wick contractions. For convenience we dress a list here of the some of three-point functions that are not trivially zero, suitably normalized.
	
    Define the normalized correlator as
	\begin{equation}
		\normCorr{O_1 \dots O_k} \equiv \frac{\expval{O_1 \dots O_k}}{\sqrt{\expval{O_1 \bar O_1}\dots \expval{O_k \bar O_k}}} \,.
	\end{equation}
	Then:
	\begin{align}
		\normCorr{(\Tr X^2) (\Tr \bar X^2) (\Tr X \bar X - \Tr Y \bar Y)} &\ = \  \sqrt{\frac{2}{N^2 - 1}} \\
		\normCorr{(\Tr XY) (\Tr \bar X\bar Y) (\Tr X \bar X + \Tr Y \bar Y - 2 \Tr Z \bar Z)} &\ = \  \sqrt{\frac{2}{3(N^2 - 1)}} \\
		\normCorr{(\Tr X^2)^2 (\Tr \bar X^2) (\Tr \bar X^2)} &\ = \   \sqrt{2\frac{N^2 + 1}{N^2 - 1}} \\
		\normCorr{(\Tr X^2)^2 (\Tr \bar X^2)^2 (\Tr X \bar X - \Tr Y \bar Y)} &\ = \  2 \sqrt{\frac{2}{N^2 - 1}} \\
		\normCorr{O_{1/4} (\Tr \bar X^2) (\Tr \bar Y^2)} &\ = \  \sqrt{\frac{2}{3} \frac{N^2 - 4}{N^2 - 1}} \\
		\normCorr{O_{1/4} \bar O_{1/4} (\Tr X \bar X + \Tr Y \bar Y - 2 \Tr Z \bar Z)} &\ = \  \sqrt{\frac{2}{3(N^2 - 1)}} \\
		\normCorr{O_{1/2} (\Tr \bar X^2) (\Tr \bar Y^2)} &\ = \  \sqrt{\frac{N^2 + 1}{3(N^2 - 1)}} \\
		\normCorr{O_{1/2} \bar O_{1/2} (\Tr X \bar X + \Tr Y \bar Y - 2 \Tr Z \bar Z)} &\ = \  \sqrt{\frac{2}{3(N^2 - 1)}} \\
		\normCorr{O_{1/4} \bar O_{1/2} (\Tr X \bar X + \Tr Y \bar Y - 2 \Tr Z \bar Z)} &\ = \  0
	\end{align}

    \subsection{Mode decomposition}

    In section~\ref{sub:holography}, one task is to extract the contribution of particular scalar harmonics in sums of the form:
    \begin{equation}
        \begin{aligned}
            S(x, y) &\equiv \sum s^{k,I}(x)\, Y^{k,I}(y) \,,\\
            T_{(ab)}(x, y) &\equiv \sum t_{s}^{k,I_1}(x)\, D_{(a}D_{b)} Y^{k,I_1}(y) + t_{v}^{k,I_5}(x)\, D_{(a} Y_{b)}^{k,I_5}(y) + t_{t}^{k,I_{14}}(x) Y_{(ab)}^{k,I_{14}}(y) \,.
        \end{aligned}
    \end{equation}
    In the first case, one can integrate against a specific scalar harmonics and use the orthogonality property of the basis of harmonics to obtain
    \begin{equation}
        s^{k, I}(s) = \frac{1}{\pi^3 z(k)} \int_{S^5} S(x,y) \overline{Y^{k,I}}(y) \,.
    \end{equation}
    The second case is slightly more involved, one first needs to isolate the scalar contributions through derivatives. Using the identity
    \begin{equation}
        D^a D_{(a} D_{b)} Y^{k, I} = 4 \qty(1 + \frac{\Lambda_k}{5}) D_b Y^{k,I} \,, 
    \end{equation} 
    one obtains
    \begin{equation}
        D^a D^b T_{(ab)} = \sum 4 \Lambda_k \qty(1 +\frac{\Lambda_k}{5}) t_{s}^{k,I_1}(x)\, Y^{k,I_1}(y) \,,
    \end{equation}
    and one can then proceed as for the first sum $S(x,y)$ to extract $t_{s}^{k,I_1}$.

    \section{Asymptotic data of the \halfBPS{} and \quartBPS{} solutions}
    \label{app:vevs}

    In section~\ref{sec:cft}, we presented the vacuum expectation values of the dimension-2 chiral primaries  \eqref{eq:vevs}, as well as the charges \eqref{eq:charges}, in terms of the asymptotic expansion of the fields at the boundary \eqref{eq:bdy_expansion}. In this appendix, we write down explicitly the asymptotic data for \halfBPS{} and \quartBPS{} solutions. In the second case, one can in general only obtain perturbative results, but in the special case where the amplitudes of the perturbations are equal, one can also obtain analytical results.

    We start with the \halfBPS{} geometries. Using \eqref{eq:half_bps_solution}, we find
    \begin{equation}
        \begin{gathered}
            \lambda_1^{(1)} = \sinh 2\beta_1 \,,\quad \mu_1^{(1)} = -2\mu_2^{(1)} = \frac{4}{3} \sinh^2 \beta_1 \,,\quad \phi_1^{(1)} = -2\sinh^2 \beta_1 \,,\\
            \lambda_2^{(1)} = \phi_2^{(1)} = 0 \,,\quad \omega_0^{(1)} = -2 \omega_1^{(1)} = \frac{1}{3} \sinh^2 \beta_1  \,.
        \end{gathered}
    \end{equation}

    For the general \quartBPS{} geometries, we use the perturbative results of section~\ref{sub:perturbations}. The computations of section~\ref{sub:holography} require the asymptotic data up to order $\epsilon^4$. We find
    {\allowdisplaybreaks
    \begin{align}
    \lambda_1^{(1)} &= 2\beta_1\,\epsilon 
                    + \left(\frac{4\beta_1^{3}}{3} - 2\beta_1\beta_2^{2}\right)\epsilon^{3}, \notag\\
    \lambda_2^{(1)} &= 2\beta_2\,\epsilon 
                    + \left(\frac{4\beta_2^{3}}{3} - 2\beta_1^{2}\beta_2\right)\epsilon^{3}, \notag\\
    \mu_1^{(1)} &= \frac{2}{3}\left(2\beta_1^{2} - \beta_2^{2}\right)\epsilon^{2}
                + \frac{2}{9}\left(2\beta_1^{4} - 4\beta_2^{2}\beta_1^{2} - \beta_2^{4}\right)\epsilon^{4}, \notag\\
    \mu_2^{(1)} &= -\frac{2}{3}\left(\beta_1^{2} - 2\beta_2^{2}\right)\epsilon^{2}
                - \frac{2}{9}\left(\beta_1^{4} + 4\beta_2^{2}\beta_1^{2} - 2\beta_2^{4}\right)\epsilon^{4}, \\
    \phi_1^{(1)} &= -2\beta_1^{2}\epsilon^{2}
                - \frac{2}{3}\left(\beta_1^{4} - 4\beta_1^{2}\beta_2^{2}\right)\epsilon^{4}, \notag\\
    \phi_2^{(1)}  &= -2\beta_2^{2}\epsilon^{2}
                - \frac{2}{3}\left(\beta_2^{4} - 4\beta_1^{2}\beta_2^{2}\right)\epsilon^{4}, \notag\\
    \omega_0^{(1)} &= \frac{1}{3}\left(\beta_1^{2} + \beta_2^{2}\right)\epsilon^{2}
                    + \frac{1}{9}\left(\beta_1^{4} - 8\beta_2^{2}\beta_1^{2} + \beta_2^{4}\right)\epsilon^{4}, \notag\\
    \omega_1^{(1)} &= -\frac{1}{6}\left(\beta_1^{2} + \beta_2^{2}\right)\epsilon^{2}
                    - \frac{1}{18}\left(\beta_1^{4} - 8\beta_2^{2}\beta_1^{2} + \beta_2^{4}\right)\epsilon^{4}. \notag
    \end{align}
    }
    In the special case when the amplitudes of both perturbations are equal, $\beta_1 = \beta_2$, one can expand the equations of motion \eqref{eq:decoupled_bps} near the boundary to obtain results in terms of the constant of motion $C$ as well as an additional function $g$ of $\beta_1=\beta_2$:
    \begin{equation}
        \begin{gathered}
        \lambda_1^{(1)} = \lambda_2^{(1)} = \sqrt{C}, \qquad
        \mu_1^{(1)} = \mu_2^{(1)} = \frac{C}{6}+g, \\
        \phi_1^{(1)} = \phi_2^{(1)} = -\frac{C}{2}-3\,g, \qquad
        \omega_0^{(1)} = \frac{C}{6}+g, \qquad \omega_1^{(1)} = -\frac{C}{12}-\frac{g}{2}\,.
        \end{gathered}
    \end{equation}
    Note that $C$ and $g(\beta_1)$ also have perturbative expansions in terms of $\beta_1$
    \begin{align}
        C =& \,4 \beta_1^2 \epsilon^2 - \frac{8}{3} \beta_1^4 \epsilon^4 + \frac{148}{45} \beta_1^6 \epsilon^6 + \dots \,,\notag\\
        g =& -\frac{2}{9}\,\beta_1^4\,\epsilon^4+\frac{8}{27}\,\beta_1^6\,\epsilon^6+\dots \,
    \end{align}
    but alternatively one can treat $C$ as the new amplitude parameter, and then one has
    \begin{equation}
        g = -\frac{1}{18} C^2 + \frac{11}{12960} C^4 + \dots
    \end{equation}

    \section{Expression of the neutral single-particle operator of dimension 4}
    \label{app:spOp4}

    In equation \eqref{eq:O40}, we define a single-particle operator of dimension 4, and further show that its expectation value is not protected. For convenience we write here the full expression of this operator in terms of the fundamental scalar fields of $\mathcal{N}=4$ SYM:
    \begin{align}
        O_{4(0)} =&~ E_1^2 E_2^2 O_{4(4)}
        \\
        =&~  16\Tr XX\bar X\bar X
            + 16\Tr ZZ\bar Z\bar Z
            + 8\Tr Z\bar Z Z\bar Z
            + 8\Tr X\bar X X\bar X \notag
        \\
        &   - 16\Tr XZ\bar X\bar Z
            - 16\Tr X\bar Z\bar X Z
            - 16\Tr XZ\bar Z\bar X \notag
        \\
        &   - 16\Tr X\bar Z Z\bar X
            - 16\Tr X\bar X Z\bar Z
            - 16\Tr X\bar X\bar Z Z
        \\
        &   - \frac {8 (2 N^2 - 3)} {N (N^2 + 1)}
            \Big[2(\Tr X\bar X - \Tr Z\bar Z)^2 + (\Tr X^2) (\Tr\bar X^2) + (\Tr Z^2) (\Tr\bar Z^2) \notag
        \\
        &\qquad\qquad - 4(\Tr X\bar Z) (\Tr\bar X Z)
            - 4 (\Tr XZ)(\Tr\bar X\bar Z)  \Big] \notag \,.
    \end{align}

	\newpage
	\bibliographystyle{jhep}
	\bibliography{biblio}

\end{document}